\documentclass[fleqn,usenatbib]{mnras}

\usepackage{newtxtext,newtxmath}

\usepackage[T1]{fontenc}
\usepackage{ae,aecompl}
\usepackage{lscape}

\usepackage{graphicx}	
\usepackage{amsmath}	
\usepackage{amssymb}	

\newcommand{\JWST}{{\it JWST}}
\newcommand{\NGTS}{NGTS}
\newcommand{\TESS}{{\it TESS}}
\newcommand{\gpe}{{GP-EBOP}}

\newcommand{\Nstar}{NGTS-1}
\newcommand{\Nkamp}{\mbox{$166 \pm 11$}}
\newcommand{\Ngamma}{\mbox{$97.1808 \pm 0.0070$}}
\newcommand{\NRA}{\mbox{$05^{\rmn{h}} 30^{\rmn{m}} 51\fs45$}} 
\newcommand{\NDec}{\mbox{$-36\degr 37\arcmin 50\farcs 83$}} 
\newcommand{\NpropRA}{\mbox{$-32.9\pm2.0 $}} 
\newcommand{\NpropDec}{\mbox{$-39.6\pm1.8$}} 
\newcommand{\Nstarmass}{\mbox{$0.617\,^{+0.023}_{-0.062}$}}
\newcommand{\Nstarradius}{\mbox{$0.573 \pm 0.077$}}
\newcommand{\Nstardensity}{\mbox{$4.4\,^{+2.4}_{-1.4}$}}
\newcommand{\Nteff}{\mbox{$3916\,^{+71}_{-63}$}}
\newcommand{\Nrotation}{\mbox{43}} 
\newcommand{\Nvmag}{\mbox{$15.52\pm 0.08$}}
\newcommand{\Nmetal}{\mbox{0}} 
\newcommand{\Ndist}{\mbox{$224\,^{+37}_{-44}$}}
\newcommand{\Nlogg}{\mbox{$4.71 \pm 0.23$}}
\newcommand{\Ngalu}{\mbox{$11.3 \pm 8.3$}}
\newcommand{\Ngalv}{\mbox{$-57.2 \pm 1.6$}}
\newcommand{\Ngalw}{\mbox{$-79.7 \pm 7.6$}}
\newcommand{\Nvsini}{$<1.0$}

\newcommand{\Nplanet}{NGTS-1b}
\newcommand{\Nperiod}{\mbox{$2.647298 \pm 0.000020$}}
\newcommand{\Nperiodshort}{\mbox{$2.647$}}
\newcommand{\Nduration}{\mbox{$1.23 \pm 0.04$}}
\newcommand{\Ntc}{\mbox{$2457720.6593965 \pm 0.00062$}}
\newcommand{\Necc}{\mbox{$0.016\,^{+0.023}_{-0.012}$}}
\newcommand{\Nmass}{\mbox{$0.812\,^{+0.066}_{-0.075}$}}
\newcommand{\Nradius}{\mbox{$1.33\,^{+0.61}_{-0.33}$}}
\newcommand{\Ndensity}{\mbox{$0.42\,^{+0.59}_{-0.28}$}}
\newcommand{\NTeq}{\mbox{$790\pm 20$}}
\newcommand{\Nflux}{\mbox{($8.89\pm0.70) \times 10^{7}$}}
\newcommand{\Nrratio}{\mbox{$0.239\,^{+0.100}_{-0.054}$}}
\newcommand{\Nau}{\mbox{$0.0326\,^{+0.0047}_{-0.0045}$}}
\newcommand{\Naoverr}{\mbox{$12.2\pm0.7$}}
\newcommand{\Nsignalsize}{\mbox{$146$}}
\newcommand{\Nimpact}{\mbox{$0.9948\,^{+0.13}_{-0.088}$}}

\newcommand{\kms}{km\,s$^{-1}$}
\newcommand{\ms}{m\,s$^{-1}$}
\newcommand{\masy}{mas\,y$^{-1}$}
\newcommand{\mpl}{\mbox{M$_{p}$}}
\newcommand{\rpl}{\mbox{R$_{p}$}}
\newcommand{\mstar}{\mbox{M$_{s}$}}
\newcommand{\rstar}{\mbox{R$_{s}$}}
\newcommand{\mjup}{\mbox{M$_{J}$}}
\newcommand{\rjup}{\mbox{R$_{J}$}}
\newcommand{\msun}{\mbox{$M_{\odot}$}}
\newcommand{\rsun}{\mbox{$R_{\odot}$}}
\newcommand{\rearth}{R$_{\oplus}$}
\newcommand{\gccc}{g\,cm$^{-3}$}
\newcommand{\ergscm}{erg\,s$^{-1}$cm$^{-2}$}
\newcommand{\teff}{$T_{\rm eff}$}
\newcommand{\logg}{$\log g$}

\newcommand{\LSO}{La Silla Observatory}

\title[\Nplanet]{\Nplanet: A hot Jupiter transiting an M-dwarf}

\author[D. Bayliss et al.]{
\parbox{\textwidth}{
Daniel~Bayliss,$^{1}$\thanks{E-mail: \href{daniel.bayliss@unige.ch}{daniel.bayliss@unige.ch}}
Edward~Gillen,$^{2}$
Philipp~Eigm\"uller,$^{3}$
James~McCormac,$^{4,5}$
~
Richard D. Alexander,$^{6}$
David J. Armstrong,$^{4,5}$
Rachel S. Booth$,^{7}$
Fran\c{c}ois Bouchy,$^{1}$
Matthew R. Burleigh,$^{6}$
Juan~Cabrera,$^{3}$
Sarah L. Casewell,$^{6}$
Alexander Chaushev,$^{6}$
Bruno Chazelas,$^{1}$
Szilard~Csizmadia,$^{3}$
Anders~Erikson,$^{3}$
Francesca~Faedi,$^{4}$
Emma~Foxell,$^{4}$
Boris~T.~G\"ansicke,$^{4,5}$
Michael R.~Goad,$^{6}$
Andrew Grange,$^{6}$
Maximilian~N.~G{\"u}nther,$^{2}$
Simon~T.~Hodgkin,$^{8}$
James Jackman,$^{4}$
James~S.~Jenkins,$^{9,10}$
Gregory Lambert,$^{2}$
Tom Louden,$^{4,5}$
Lionel Metrailler,$^{1}$
Maximiliano~Moyano,$^{11}$
Don~Pollacco,$^{4,5}$
Katja Poppenhaeger,$^{7}$
Didier~Queloz,$^{2}$
Roberto~Raddi,$^{4}$
Heike~Rauer,$^{3,12}$
Liam Raynard,$^{6}$
Alexis~M.~S.~Smith,$^{3}$
Maritza~Soto,$^{9}$
Andrew~P.~G.~Thompson,$^{7}$
Ruth~Titz-Weider,$^{3}$
St\'{e}phane~Udry,$^{1}$
Simon.~R.~Walker,$^{4}$
Christopher~A.~Watson,$^{7}$
Richard~G.~West,$^{4,5}$
Peter~J.~Wheatley$^{4,5}$
}
\\
$^{1}$Observatoire de Gen{\`e}ve, Universit{\'e} de Gen{\`e}ve, 51 Ch. des Maillettes, 1290 Sauverny, Switzerland\\
$^{2}$Astrophysics Group, Cavendish Laboratory, J.J. Thomson Avenue, Cambridge CB3 0HE, UK\\
$^{3}$Institute of Planetary Research, German Aerospace Center, Rutherfordstrasse 2, 12489 Berlin, Germany\\
$^{4}$Dept.\ of Physics, University of Warwick, Gibbet Hill Road, Coventry CV4 7AL, UK\\
$^{5}$Centre for Exoplanets and Habitability, University of Warwick, Gibbet Hill Road, Coventry CV4 7AL, UK\\
$^{6}$Department of Physics and Astronomy, Leicester Institute of Space and Earth Observation, University of Leicester, LE1 7RH, UK\\
$^{7}$Astrophysics Research Centre, School of Mathematics and Physics, Queen's University Belfast, BT7 1NN Belfast, UK\\
$^{8}$Institute of Astronomy, University of Cambridge, Madingley Rise, Cambridge CB3 0HA, UK\\
$^{9}$Departamento de Astronomia, Universidad de Chile, Casilla 36-D, Santiago, Chile\\
$^{10}$ Centro de Astrof\'isica y Tecnolog\'ias Afines (CATA), Casilla 36-D, Santiago, Chile.\\
$^{11}$Instituto de Astronomia, Universidad Cat\'{o}lica del Norte, Casa Central, Angamos 0610, Antofagasta, Chile\\
$^{12}$Center for Astronomy and Astrophysics, TU Berlin, Hardenbergstr. 36, D-10623 Berlin, Germany
}

\date{Accepted for publication on 20 October 2017}

\pubyear{2017}

\begin{document}
\label{firstpage}
\pagerange{\pageref{firstpage}--\pageref{lastpage}}
\maketitle

\begin{abstract}
We present the discovery of \Nplanet, a hot-Jupiter transiting an early M-dwarf host ($T_\mathrm{eff,*}$=\Nteff\,K) in a $P=\Nperiodshort$\,d orbit discovered as part of the Next Generation Transit Survey (\NGTS).  The planet has a mass of \Nmass\,\mjup\, making it the most massive planet ever discovered transiting an M-dwarf.  The radius of the planet is \Nradius\,\rjup.  Since the transit is grazing, we determine this radius by modelling the data and placing a prior on the density from the population of known gas giant planets. \Nplanet{} is the third transiting giant planet found around an M-dwarf, reinforcing the notion that close-in gas giants can form and migrate similar to the known population of hot Jupiters around solar type stars.  The host star shows no signs of activity, and the kinematics hint at the star being from the thick disk population.  With a deep (2.5\%) transit around a $K=11.9$ host, \Nplanet{} will be a strong candidate to probe giant planet composition around M-dwarfs via \JWST{} transmission spectroscopy.
\end{abstract}




\section{Introduction}
\label{sec:intro}
M-dwarf stars as planetary hosts are of high interest.  Two important recent discoveries in the field of exoplanets relate to planets orbiting M-dwarfs:  Proxima Centauri \citep{2016Natur.536..437A} and Trappist-1 \citep{2017Natur.542..456G}.  The interest primarily derives from the fact that the radius ratio ($R_\mathrm{P}/R_\mathrm{star}$) and mass ratio ($M_\mathrm{P}/M_\mathrm{star}$) for such systems are much higher than equivalent systems with solar-type hosts, thus they can be easier to detect via transits and radial velocities.  The low intrinsic luminosity of M-dwarfs also means that the habitable zone is very close to the host star and therefore it is much easier to detect potential habitable planets around these stars, compared to their more massive counterparts.  Finally, M-dwarfs are the most populous stars in the Galaxy \citep[$\sim$75\%;][]{henry06}, and hence understanding planet formation and planet frequency around these low mass stars greatly enhances our knowledge of the full population of planets in the Galaxy.

However, observationally M-dwarfs present some significant drawbacks when searching for exoplanets.  Firstly, the intrinsically low luminosity of M-dwarfs means that  on average they have much fainter apparent magnitudes than solar type stars, making it hard to obtain high precision photometry or high precision radial velocity measurements.  Secondly, the spectra of M-dwarfs are dominated by molecular lines which do not allow the same precision radial velocity measurements as are afforded by the unblended, sharp metal lines that can be found in solar-type spectra.  Finally the low effective temperatures of M-dwarfs means they have a peak flux well outside the optical wavelength range where most photometry and spectroscopy is targeted.

Despite these difficulties there has been progress in identifying exoplanets orbiting M-dwarfs using dedicated facilities.  Notably the Earth-sized GJ1132-b \citep{BertaThompson2015} and super-Earth GJ1214-b \citep{Charbonneau2009}, detected with the MEarth ground-based transit survey \citep{Nutzman2008}, have been popular targets for atmospheric characterisation owing to the  small radii and close proximity to Earth of their M-dwarf hosts.  A large radial velocity program to monitor M-dwarfs, carried out by the HARPS consortium \citep{Mayor2003,Bonfils2005}, has produced more than 10 confirmed exoplanets and recently two planets orbiting GJ 273 - potentially making this the closest known system with planets in the habitable zone after Proxima Centauri \citep{berdinas2016,Astudillo2017}.  The Kepler and K2 missions, while not designed to specifically target M-dwarfs, have discovered transiting planets around M-dwarfs in their FOV: Kepler-138b \citep{Rowe2014}; K2-33b \citep{David2016}.  In addition to these programs the wide-field Next Generation Transit Survey (\NGTS) survey is under-way. \NGTS{} is optimised to detect transits around K and early M-dwarfs and is at the limit of photometric precision possible for ground-based observations \citep{project2017,McCormac2017,2013EPJWC..4713002W,Chazelas2012}.

Small planets (0.5--4 \rearth) appear to be common around M-dwarfs \citep{dressing13}, and gas giants are considerably rarer.  \citet{meyer17} give the frequency of planets with masses 1-10\,\mjup\ as 0.07 planets per M-dwarf.  To date only two gas giant planets have been discovered transiting M-dwarfs - namely Kepler-45b \citep[\mpl=0.505\,\mjup, \rpl=0.96\,\rjup][]{johnson12} and HATS-6b \citep[\mpl=0.32\,\mjup, \rpl=1.00\,\rjup][]{hartman15}.  WASP-80b  \citep[\mpl=0.55\,\mjup, \rpl=0.95\,\rjup][]{wasp80} also transits a low mass host star very similar to these early M-dwarfs, and is close in its physical properties to Kepler-45b.  However hot Jupiters are rare around solar type stars also - from Kepler statistics the frequency of hot-Jupiter systems around solar-type stars is $0.43\pm0.05$ for $P<10$\,d \citep{fressin13}.  Coupled to this, transit surveys have not monitored large numbers of M-dwarfs compared with the solar-type population.  Thus a larger population of M-dwarfs needs to be monitored in order to ascertain if the frequency of hot-Jupiters around M-dwarfs significantly departs from the solar-type frequency.  

In this paper we report the discovery of \Nplanet, the third gas giant found to transit an M-dwarf.  In Section~\ref{sec:obs} we detail the observations that led to the discovery of this planet, including the \NGTS{} survey photometry, high precision follow-up photometry, and high resolution spectroscopy.  In Section~\ref{sec:analysis} we analyse the observational data in conjunction with archival photometric data to precisely determine the physical characteristics of \Nplanet.  Finally, in Section~\ref{sec:discussion} we discuss the results including the significance of this discovery in terms of planet formation scenarios around low mass stars and the prospects for characterisation via JWST. 


\section{Observations}
\label{sec:obs}
The discovery of \Nplanet{} was made using the \NGTS{} telescopes in conjunction with high precision photometric follow-up from Eulercam and high precision spectroscopy from HARPS.  We detail all of these observations in this Section.

\subsection{\NGTS{} Photometry}
\label{sub:ngtsphot}
\NGTS{} is a fully automated array of twelve 20\,cm aperture Newtonian telescopes situated at the ESO Paranal Observatory in Chile.  Each telescope is coupled to an Andor Ikon-L Camera featuring a 2K$\times$2K e2V deep-depleted CCD with 13.5\,$\mu$m pixels, which corresponds to an on-sky size of 4.97\,\arcsec.  Telescopes are operated from independent Optical Mechanics, Inc. (OMI) mounts, allowing for excellent tracking and freedom in scheduling observations.

\Nstar\ was observed on a single \NGTS{} telescope over a photometric campaign conducted between 2016 August 10 and 2016 December 7.  In total we obtained 118,498$\times$10\,s exposures in the \NGTS{} bandpass (550 -- 927\,nm) over $\sim$100 usable nights.  Raw data were reduced to calibrated fits frames via the \NGTS{} reduction pipeline fully described in \citet{project2017}.  Photometry was extracted via aperture photometry using the \NGTS{} photometric pipeline, which is a customised implementation of the CASUTools\footnote{\url{http://casu.ast.cam.ac.uk/surveys-projects/software-release}} photometry package.  To account for systematic effects in the data, an implementation of the SysRem algorithm \citep{Tamuz2005} was applied to the data.  Again full details can be found in \citet{project2017}.

Light curves were searched for transit-like signals using ORION, an implementation of the box-fitting least squares (BLS) algorithm \citep{Kovacs2002}.  We identified a strong BLS signal for \Nstar{} at a period of \Nperiod\,d.  

\begin{figure}
	\includegraphics[width=\columnwidth]{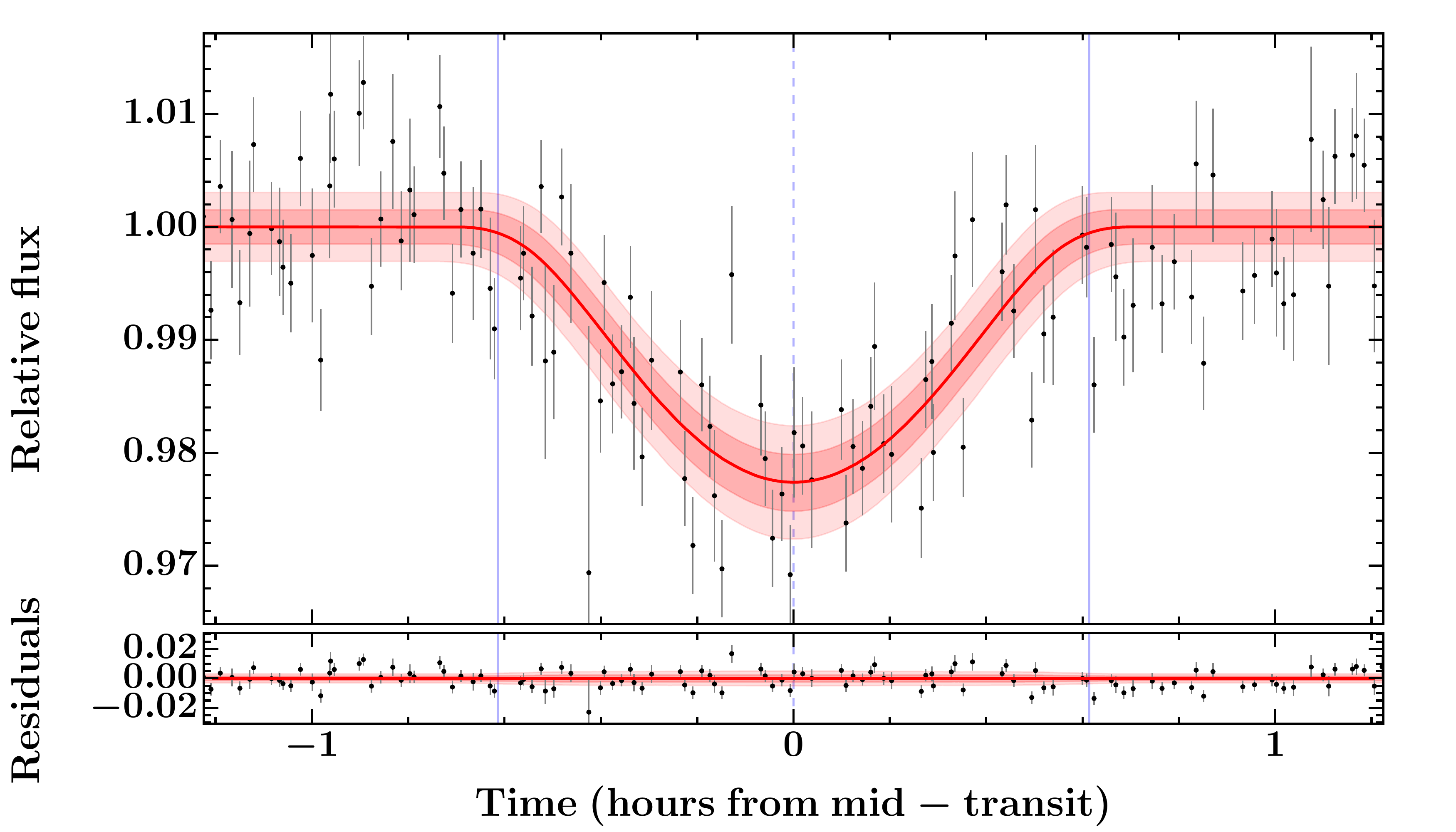}
    \caption{The \NGTS\ light curve of \Nstar\, phase-folded at the best-fitting period of \Nperiod\,d and zoomed to show the transit event.  Black circles show the photometric datapoints binned to 10-minutes with associated photon error.  The red line and pink shaded regions show the median and the 1 and 2-sigma uncertainties, respectively, of the posterior model based on the global modelling set out in Section~\ref{sub:global}.  Solid blue vertical lines indicate the start of ingress (T$_{1}$) and end of egress (T$_{4}$).  Dashed blue vertical line represents the transit centre (T$_{\rm c}$).  Residuals to the global model are shown below the light curve.} 
    \label{fig:ngtsphot}
\end{figure}

We set out the transit feature identified in the \NGTS{} light curve, phase-folded to this period, in Figure~\ref{fig:ngtsphot}, and provide the full light curve dataset in Table~\ref{tab:ngts}. 

\begin{table}
	\centering
	\caption{\NGTS{} photometry for \Nstar.  The full table is available in a machine-readable format from the online journal.  A portion is shown here for guidance.}
	\label{tab:ngts}
	\begin{tabular}{ccc}
	Time	&	Flux        	&Flux\\
    (HJD-2450000)	&	(normalised)	&error\\
	\hline
    7610.84860  &  0.9633  &  0.0315 \\
	7610.84875  &  1.0530  &  0.0317 \\
	7610.84890  &  1.0636  &  0.0317 \\
	7610.84905  &  1.0216  &  0.0317 \\
	7610.84920  &  0.9808  &  0.0317 \\
	7610.84935  &  1.0042  &  0.0317 \\
	7610.84949  &  1.0401  &  0.0317 \\
	7610.84964  &  1.0316  &  0.0316 \\
	7610.84979  &  0.9792  &  0.0316 \\
	7610.84994  &  1.0067  &  0.0317 \\
        ...        &   ...    &   ...   \\
	\hline
	\end{tabular}
\end{table}

Our previous yield simulations have evaluated the risk of false positives mimicking a planetary transit such as \Nstar\ \citep{Guenther17a}.  In order to verify that this transit signal is consistent with a transiting planet, we applied a series of vetting tests to the \NGTS\ data. 

Firstly, we checked the light curve for evidence of a secondary eclipse near phase 0.5.  Such a detection, at our photometric precision, would indicate an eclipsing stellar companion rather than a transiting planet companion.  However no secondary eclipse is detected.

Secondly we checked the odd/even transit depth differences to help rule out the possibility that we have detected half the true period, and have phase-folded primary and secondary eclipses together.  We see no odd/even depth difference to the precision of our photometry.

Thirdly we check for out-of-transit ellipsoidal variation that is present in many short-period binary stars.  Again, no such variability is seen to the precision of our photometry. 

Finally we check for a shift in the photometric centre-of-flux during transit events as fully described in \cite{Guenther17b}.  This test is able to detect blended eclipsing binaries at separations well below the size of individual \NGTS\ pixels - i.e., to the level of $\sim$1\arcsec.  We find no centroiding variation during the transit events of \Nstar. 

Using existing photometric and proper motion survey data (see Table \ref{tab:stellar}), we identified that \Nstar{} is a low temperature star, and in combination with the stellar populations from the Besan\c{c}on Galaxy model \citep{robin03} we concluded that \Nstar{} is most likely an M-dwarf rather than an M-giant.  This means that the deep  (2.5~\%) and  V-shaped transit is still consistent with a transiting body being of planetary radius, and therefore we passed this candidate on for further photometric and spectroscopic follow-up, as detailed in Sections \ref{sub:eulerphot} and \ref{sub:spect}.


\subsection{Eulercam Photometry}
\label{sub:eulerphot}
\begin{figure}
	\includegraphics[width=\columnwidth]{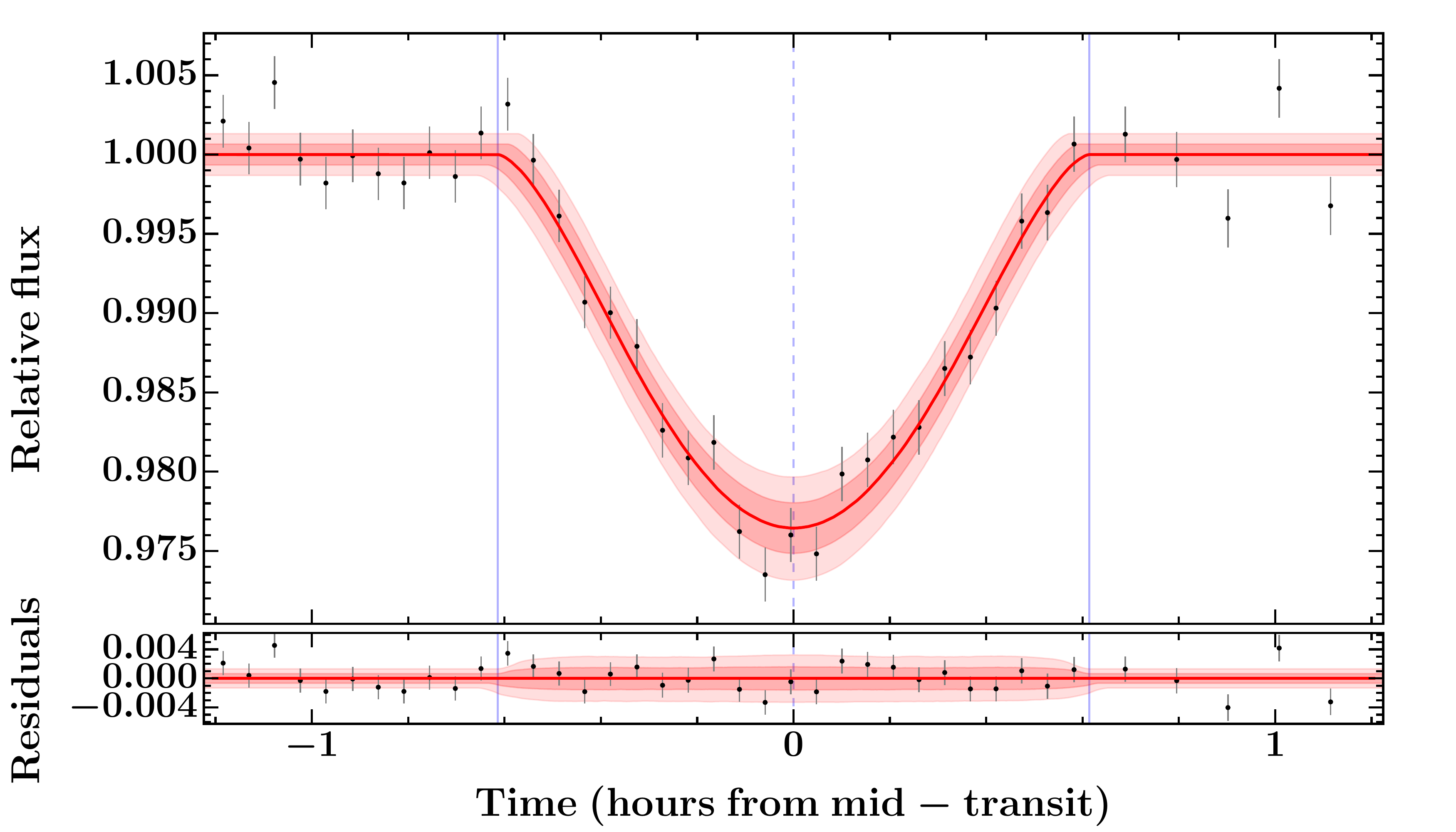}
    \caption{The z-band Eulercam light curve for \Nstar\ on the UT night of 2017 February 26 from \LSO, Chile.  Black datapoints are the individual 180\,s exposures.   All other symbols as for Figure~\ref{fig:ngtsphot}.}
    \label{fig:eulercam}
\end{figure}

In order to refine the ephemeris, increase the precision of the light curve component of the global modelling (see Section~\ref{sub:global}), and check for any evidence of a blend via a colour-dependent depth difference, we obtained a high precision light curve for \Nstar\, using Eulercam on the 1.2\,m Euler Telescope at \LSO.  Observations were taken on the night of 2017 February 26 between 2:14 and 5:08UT in the z-band with the telescope slightly defocused (0.5mm) to broaden the point-spread function.  In total 50 exposures were obtained, each with an exposure time of 180\,s.  After the images were reduced via standard bias and flatfield correction, we performed aperture photometry on \Nstar{} and a set of ten reference stars using a photometric pipeline utilising Source Extractor \citep{1996A&AS..117..393B}.  A full transit event was observed, and the photometric data are presented in Table~\ref{tab:eulercam}, while the data are plotted with a best fit transit model in Figure~\ref{fig:eulercam}.  These data are used in the global modelling for the \Nplanet{} system detailed in \ref{sub:global}.      

\begin{table}
	\centering
	\caption{Z-band Eulercam photometry for \Nstar.  The full table is available in a machine-readable format from the online journal.  A portion is shown here for guidance.}
	\label{tab:eulercam}
	\begin{tabular}{ccc} 
	Time	&	Flux        	&Flux\\
    (HJD-2450000)	&	(normalised)	&error\\
	\hline
7810.59349558& 1.000691& 0.001658\\
7810.59571697& 1.001244& 0.001659\\
7810.59793916& 1.000230& 0.001657\\
7810.60016845& 0.999125& 0.001655\\
7810.60247724& 1.001336& 0.001659\\
7810.60469833& 1.000415& 0.001657\\
7810.60691832& 1.004940& 0.001665\\
7810.60921191& 0.999217& 0.001655\\
7810.61142560& 1.002259& 0.001660\\
7810.61364589& 1.000875& 0.001658\\
...& ...& ...\\
	\hline
	\end{tabular}
\end{table}

\subsection{Spectroscopy}
\label{sub:spect}
We obtained multi-epoch spectroscopy for \Nstar\ with the HARPS spectrograph \citep{2003Msngr.114...20M} on the ESO 3.6\,m telescope at \LSO, Chile, between 2017 January 27 and 2017 February 7.  Due to the faint optical magnitude of \Nstar{} (V=\Nvmag), we used the HARPS in the high efficiency fibre link (EGGS) mode, which utilises a slightly larger fibre	(1.4\arcsec compared with the 1.0\arcsec\, in nominal mode).  This gives a higher signal-to-noise spectrum, albeit with a slightly lower resolution (R=85000 compared with R=110000).  To further increase the signal-to-noise we took long exposures (1-hour) for each epoch.  

\begin{figure}
	\includegraphics[width=\columnwidth]{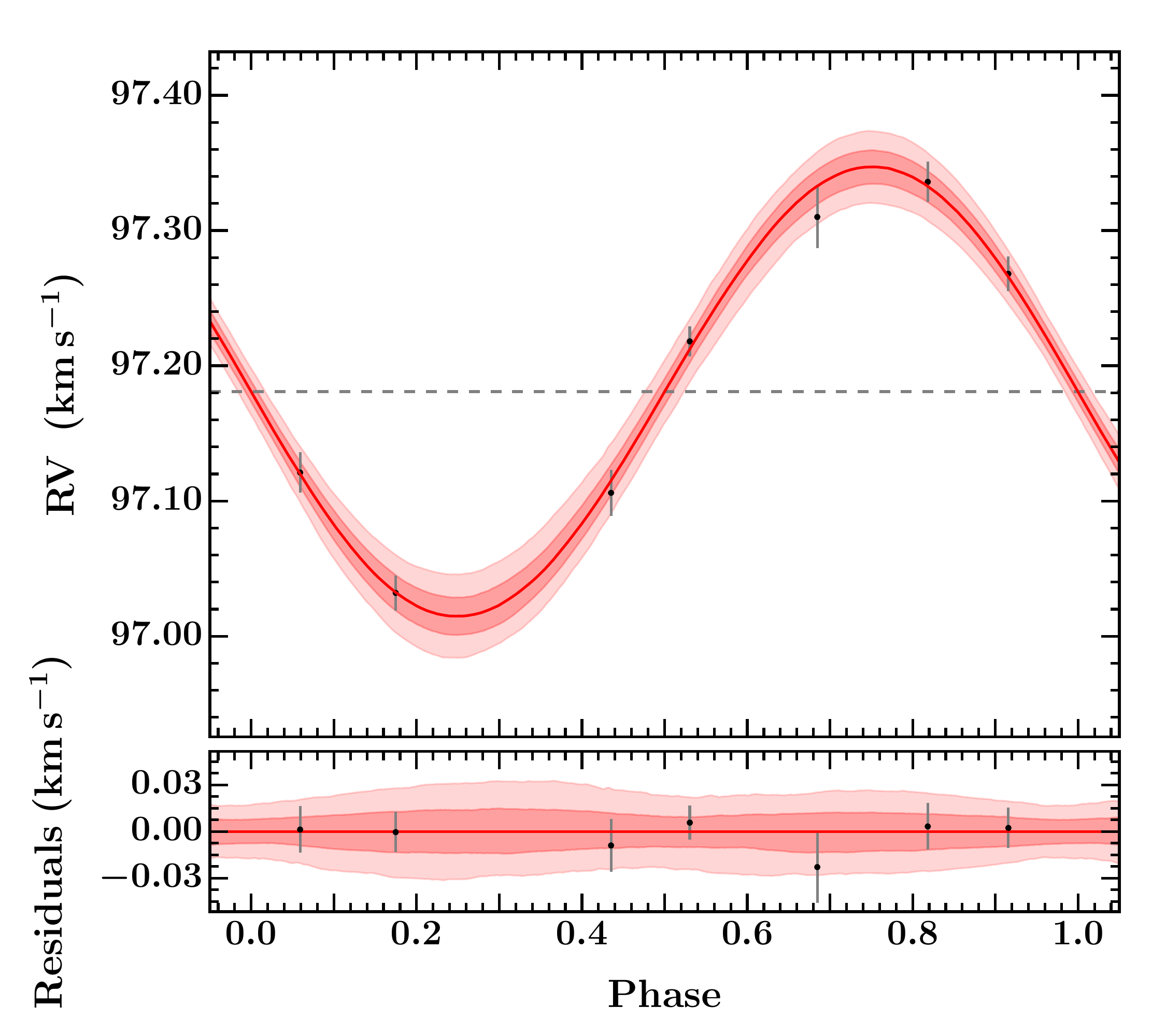}
    \caption{Phase-wrapped radial velocity measurements for \Nstar{} from the HARPS spectrograph on the ESO 3.6\,m at \LSO, Chile.  The red line and pink shaded regions show the median and the 1 and 2-sigma uncertainties, respectively, of the posterior model based on the global modelling set out in Section~\ref{sub:global}. 
The horizontal dashed line shows the systemic radial velocity of \Nstar.  The residuals to the global model are shown below the radial velocity data.}
    \label{fig:harps}
\end{figure}

Using the standard HARPS data reduction pipeline, radial velocities were calculated for each epoch via cross-correlation with the M2 binary mask.  These radial velocities are listed, along with their associated error, FWHM, contrast, and bisector slope, in Table~\ref{tab:rvs}.  The radial velocities show a variation in-phase with the photometric period and phase.  The semi-amplitude is large ($K=$\Nkamp \,\ms), indicating a Jupiter-mass planet.  We plot the phase-wrapped radial velocities in Figure~\ref{fig:harps}.  Again, these data are used in the global modelling for the \Nplanet{} system detailed in Section~\ref{sub:global}.  

We also wavelength shift and combine all seven spectra to create a single, high signal-to-noise spectrum from HARPS.  This spectrum is used to characterise the stellar properties of \Nstar; see Section~\ref{sub:stellar}.

\begin{table*}
	\centering
	\caption{Radial Velocities for \Nstar}
	\label{tab:rvs}
	\begin{tabular}{cccccc} 
HJD			&	RV		&RV err &	FWHM& 	contrast&BIS\\
(-2450000)	& (\kms)& (\kms)&(\kms) & &(\kms) \\
		\hline
7780.712430&	97.310&	0.023&	3.783&	10.7&	-0.080\\
7784.657148&	97.032&	0.013&	3.718&	11.3&	-0.011\\
7785.598834&	97.218&	0.011&	3.730&	11.3&	-0.011\\
7786.618897&	97.268&	0.013&	3.671&	11.1&	-0.044\\
7789.645749&	97.121&	0.015&	3.737&	11.1&	-0.035\\
7790.641250&	97.106&	0.017&	3.735&	10.7&	-0.012\\
7791.655373&	97.336&	0.015&	3.693&	10.5&	-0.006\\
		\hline
	\end{tabular}
\end{table*}

\section{Analysis}
\label{sec:analysis}
The observations made in Section~\ref{sec:obs} are combined with archival data from various photometric surveys in order to fully characterise the \Nplanet{} system.  In this Section we detail that analysis, in addition to analysing the data for stellar activity, rotation, and background contamination.


\subsection{Stellar Properties}
\label{sub:stellar}
One of the difficulties with characterising M-dwarfs is that in the optical they are faint and have spectra dominated by strong molecular bands \citep{lepine13}.  Metallicity is especially difficult to determine, with the best methods requiring K-band spectra to measure the CaI and NaI lines from 2.2--2.3\,$\mu$m \citep{rojas10}.  This impacts on the stellar parameters since the stellar radius for M-dwarfs is highly metallicity dependent.

Nevertheless, we are able to use the combined HARPS spectrum (see Section~\ref{sub:spect}) to spectral-type \Nstar.  We compared the combined HARPS spectrum using M-dwarf spectral templates derived from the Sloan Digital Sky Survey \citep{rebassa-mansergasetal07-1}. The HARPS spectrum is not globally flux-calibrated, and we therefore fitted the M-dwarf templates over a 100\,\AA\ wide band of the HARPS spectrum, recording the value of the local $\chi^2$. This band was then sliced across the full spectral range, and a global $\chi^2$ was computed from the sum of all the local values. The spectral type determined from this procedure is M$0.5\pm0.5$, any later types are clearly excluded by the relative strengths of the molecular absorption bands, and by the width of the Na\,I 5890/96\,\AA\ lines.  Although metallicity cannot be well determined, we see no evidence of the star being extremely metal rich or poor in comparison to M-dwarf templates \citep{savcheva2014}, and therefore we adopt a metallicity of $[{\rm M/H}]=0$ for our analysis.  From the HARPS spectrum we can also place an upper limit on the $v$sin$i$ of \Nstar\ at 1.0\,\kms, consistent with the possible rotational period derived from the photometry and discussed in Section~\ref{sub:activity}.

\begin{figure}
	\includegraphics[width=\columnwidth]{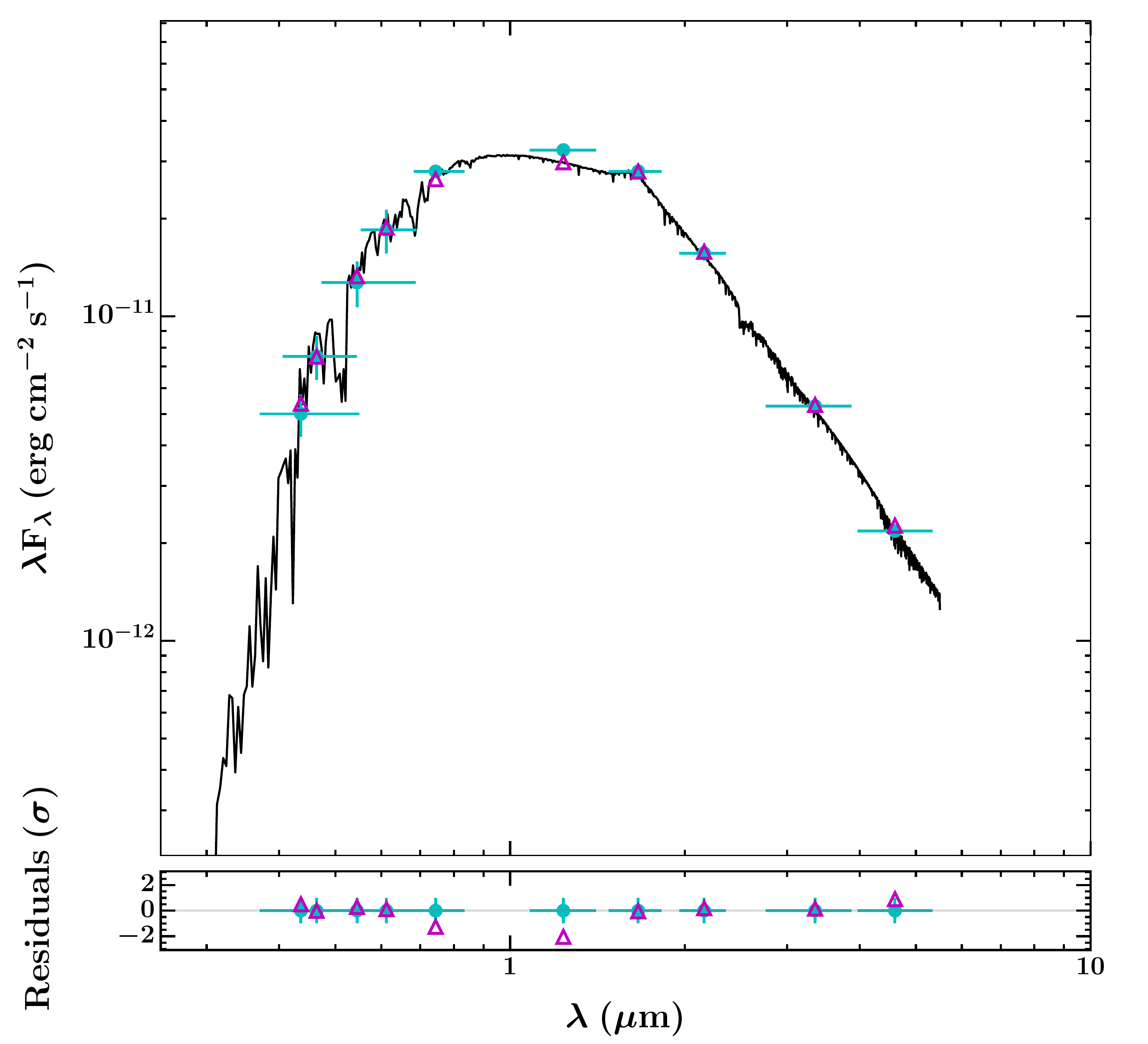}
    \caption{The fitted spectral energy distribution (black line) for \Nstar\ based on the photometric data (cyan points) presented in Table~\ref{tab:stellar}. The magenta triangles show the model flux at the wavelengths of the photometric data.}
    \label{fig:sed}
\end{figure}

Given the indication from the combined HARPS spectrum that the host was an early M-dwarf, we utilise the recent empirical relations that are based on extensive studies of nearby M-dwarfs.   In particular, we use the relations of \citet{Mann15} (hereafter M15) and \citet{Benedict16} (hereafter B16), along with modelling the stellar SED, to estimate mass, radius and \teff\ values for \Nstar. We took three main steps: 
\begin{enumerate}

\item We estimated \teff, radius and mass using M15. \@ \teff\ was estimated from the system's broadband colours, which gave $T_{\rm eff} = 3867\pm107$\,K. The radius was then estimated from this \teff\ value, which gave $\rstar=0.573 \pm 0.077\,\rsun$\ (note this is not a tight relation given we do not know the metallicity). The radius was then used to determine the absolute $K_{s}$ band magnitude ($M_{K_{s}}$), which in turn gave an expected mass of $\mstar=0.598\pm0.011\,\msun$. This step is not particularly intuitive, but it becomes possible because of the implicit \teff\ prediction underlying the radius estimation and the fact that the error ellipses in \teff--radius space and radius--$M_{K_{s}}$ space are orthogonal. We use this mass estimate only to check our final value. 

\item We modelled the spectral energy distribution (SED) of \Nstar\ (Figure~\ref{fig:sed}) using the method described in \citet{Gillen17}. Briefly, we convolved the BT-SETTL and PHOENIX v2 model atmospheres \citep{Allard2011} with the available bandpasses (see Table \ref{tab:stellar}) and interpolated in \teff--\logg\ space (fixing [M/H]=0) within a Markov chain Monte Carlo (MCMC) and placed the following priors: a uniform (but not uninformative) prior on the radius: $0.4\,\rsun \leq \rstar \leq 0.7\,\rsun$, which brackets plausible radius estimates given the expected \teff\ range; a Gaussian prior on \logg: $\log g = 4.7 \pm 0.2$, which is estimated from the predictions of the PARSEC v1.2 stellar evolution models given our \teff, $M$ and $R$ estimates (with the uncertainty further inflated); a Gaussian prior on $A_{V}$: $A_{V} = 0.0 \pm 0.1$ (positive values allowed only) following initial tests implying a low-to-negligible reddening; uninformative priors on the \teff\ and distance; and an uninformative prior on the uncertainties, which are fit for within the MCMC. Combining the results from the BT-SETTL and PHOENIX models gives $T_{\rm eff} =$ \Nteff\,K and distance $d =$ \Ndist\,pc.

\item We then used our SED modelling results and the $M_{K_{s}}$--\,mass relation of B16 to determine $\mstar =$ \Nstarmass\,\msun. We note that this is in agreement with the M15 value.

\end{enumerate}

We adopt, as our final stellar parameters, the \teff\ and distance estimates from our SED modelling, the radius estimate from M15, and the mass estimate from B16; these are reported in Table \ref{tab:stellar}.

\begin{table}
	\centering
	\caption{Stellar Properties for \Nstar}
	\begin{tabular}{lcc} 
	Property	&	Value		&Source\\
	\hline
    \multicolumn{3}{l}{Astrometric Properties}\\
    R.A.		&	\NRA			&2MASS	\\
	Dec			&	\NDec			&2MASS	\\
    2MASS I.D.	& J05305145-3637508	&2MASS	\\
    $\mu_{{\rm R.A.}}$ (\masy) & \NpropRA & UCAC4 \\
	$\mu_{{\rm Dec.}}$ (\masy) & \NpropDec & UCAC4 \\
    \\
    \multicolumn{3}{l}{Photometric Properties}\\
	V (mag)		&$15.524 \pm 0.083$		&APASS\\
	B (mag)		&$16.912 \pm 0.077$		&APASS\\
	g (mag)		&$16.288 \pm 0.000$		&APASS\\
	r (mag)		&$14.999 \pm 0.006$		&APASS\\
	i (mag)		&$14.328 \pm 0.022$		&APASS\\
    G$_{GAIA}(mag)$	&$14.702 \pm 0.001$		&{\em Gaia}\\
    NGTS (mag)	&$14.320 \pm 0.02$		&This work\\
    J (mag)		&$12.702\pm0.024$		&2MASS	\\
   	H (mag)		&$12.058\pm0.022$		&2MASS	\\
	K (mag)		&$11.936\pm0.027$		&2MASS	\\
    W1 (mag)	&$11.800\pm0.023$		&WISE	\\
    W2 (mag)	&$11.779\pm0.022$		&WISE	\\
    \\
    \multicolumn{3}{l}{Derived Properties}\\

    T$_{\rm eff}$ (K)    & \Nteff               &SED fitting\\
    $\left[M/H\right]$			     & \Nmetal					    &HARPS Spectrum\\
    vsini (\kms)	     &	\Nvsini			    &HARPS Spectrum\\
    $\gamma_{RV}$ (\kms) & \Ngamma		        &HARPS Spectrum\\
    log g                &		\Nlogg			&ER\\
    \mstar (\msun) & \Nstarmass		        &ER\\
    \rstar (\rsun) & \Nstarradius	            &ER\\
    $\rho$ (\gccc) & \Nstardensity              & ER\\
    Distance (pc)	&  \Ndist	                &SED fitting\\
    
	\hline
    \multicolumn{3}{l}{2MASS \citep{2MASS}; UCAC4 \citep{UCAC};}\\
    \multicolumn{3}{l}{APASS \citep{APASS}; WISE \citep{WISE};}\\
    \multicolumn{3}{l}{{\em Gaia} \citep{GAIA}}\\
    \multicolumn{3}{l}{ER = empirical relations using \citet{Benedict16}}\\
    \multicolumn{3}{l}{ and \citet{Mann15}}
	\end{tabular}
    \label{tab:stellar}
\end{table}

\subsection{Global Modelling}
\label{sub:global}
Using the photometric and radial velocity data set out in Section~\ref{sec:obs}, in addition to the stellar parameters set out in Section~\ref{sub:stellar}, we globally model the system in order to determine the stellar and planetary parameters for the \Nplanet{} system.  We use the \gpe\ transiting planet and eclipsing binary model. \gpe\ is fully described in \citet{Gillen17}, where it is applied to eclipsing binaries and a transiting brown dwarf, and we further note its tested ability to model transiting planets, e.g. \citet{pepper17}.  In \gpe, the full light curve is modelled as a Gaussian process, whose mean function is a transit model. The radial velocity data are modelled simultaneously with the available photometry and incorporates a jitter term to allow for stellar activity variations and instrumental systematics in the data.  \gpe\ explores the posterior parameter space using an affine-invariant MCMC.  We use priors on stellar density and limb darkening parameters following \citep{Gillen17}. 

One difficulty in modelling this system is that the transit is grazing, which is relatively likely given the radius ratio between the planet and the star (\Nrratio{}).  Such a configuration results in a strong degeneracy between the impact parameter ($b$) and the planetary radius (\rpl).  To help break this degeneracy we construct a prior on the radius of the planet based on the existing population of hot Jupiters.  To a first approximation the two measurable quantities that influence the radius of hot Jupiters are mass \citep{hatzes15} and incident flux \citep{demory11}.  Using the NASA Exoplanet Archive\footnote{\url{exoplanetarchive.ipac.caltech.edu}} \citep{akeson13}, we take the population of transiting hot Jupiters (\mpl$>0.2$\mjup, P$<50$d) with incident fluxes in a similar regime to \Nplanet{} ($<2\times10^{8}$\,\ergscm).  This cut on incident flux is important as it excludes highly inflated hot Jupiters at high incident fluxes \citep{demory11}.  The resulting population shows a log-normal distribution of densities, with $\ln$(density) having a mean at $-$0.29 and a standard deviation of 0.89.  We apply this log-normal density distribution as a prior on the density of \Nplanet{} during the global modelling.  We present the fitted and derived parameters using this prior as our final parameters and list them in Table \ref{tab:gp}.  We also present in Table \ref{tab:gp} the parameters without any  prior on the planet density.  

\begin{table*}  
 \centering  
 \small  
 \caption[Model parameters]{Fitted and derived parameters of the models applied to $\rm{ecc}~+~\rho_{*}~\rm{prior}$ and $\rm{ecc}~+~\rho_{*}~\&~\rho_{\rm pl}~\rm{priors}$.} 
 \label{lc_model_tab}  
 \begin{tabular}{l l l c c }  
 \noalign{\smallskip} \noalign{\smallskip} \hline  \hline \noalign{\smallskip}  
 Parameter  &   Symbol  &  Unit  & \multicolumn{2}{c}{Value} \\   
     &        &            &    \,\,\,\,\,$\rm{ecc}~+~\rho_{*}~\rm{prior}$    &    \,\,\,\,\,$\rm{ecc}~+~\rho_{*}~\&~\rho_{\rm pl}~\rm{priors}$ \\ 
 \noalign{\smallskip} \hline \noalign{\smallskip} \noalign{\smallskip}  
 \multicolumn{5}{c}{\emph{Transit parameters}} \\   
 \noalign{\smallskip} \noalign{\smallskip}   
 Sum of radii    &    $(R_{\rm{s}} + R_{\rm{p}})/ a$    &        &    $\,\,\,\,\,0.139 \pm 0.034$    &    $\,\,\,\,\,0.1021\,^{+0.0108}_{-0.0087}$    \\  [1.1ex]  
 Radius ratio    &    $R_{\rm{p}} / \rstar$    &       &    $\,\,\,\,\,0.76\,^{+0.68}_{-0.47}$    &    $\,\,\,\,\,0.239\,^{+0.100}_{-0.054}$             \\  [1.1ex]  
 Orbital inclination & $i$ & $^{\circ}$        &    $\,\,\,\,\,82.8 \pm 2.3$    &    $\,\,\,\,\,85.27\,^{+0.61}_{-0.73}$    \\  [1.1ex]  
 Orbital period    &    $P$    &    days         &    $\,\,\,\,\,2.647297 \pm 0.000021$    &    $\,\,\,\,\,2.647298 \pm 0.000020$    \\  [1.1ex]  
 Time of transit centre    &    $T_{\rm{centre}}$    &    BJD         &    $\,\,\,\,\,2457720.65939 \pm 0.00067$    &    $\,\,\,\,\,2457720.65940 \pm 0.00062$    \\  [1.1ex]  
     &    $\sqrt{e} \cos \omega$    &         &    $\,\,\,\,\,0.01 \pm 0.11$    &    $-0.001 \pm 0.098$    \\  [1.1ex]  
     &    $\sqrt{e} \sin \omega$    &         &    $\,\,\,\,\,0.05 \pm 0.12$    &    $\,\,\,\,\,0.02 \pm 0.12$    \\  [1.1ex]  
 \noalign{\smallskip} \noalign{\smallskip}Linear LDC    &    $u_{\rm{NGTS}}$    &         &    $\,\,\,\,\,0.40 \pm 0.14$    &    $\,\,\,\,\,0.46 \pm 0.13$    \\  [1.1ex]  
 Non-linear LDC    &    $u'_{\rm{NGTS}}$    &         &    $\,\,\,\,\,0.14 \pm 0.23$    &    $\,\,\,\,\,0.12 \pm 0.2$    \\  [1.1ex]  
 Linear LDC    &    $u_{\rm{Euler}}$    &         &    $\,\,\,\,\,0.30 \pm 0.11$    &    $\,\,\,\,\,0.322 \pm 0.083$    \\  [1.1ex]  
 Non-linear LDC    &    $u'_{\rm{Euler}}$    &         &    $\,\,\,\,\,0.21 \pm 0.24$    &    $\,\,\,\,\,0.21 \pm 0.18$    \\  [1.1ex]  
 \noalign{\smallskip} \noalign{\smallskip} \noalign{\smallskip} \noalign{\smallskip}  \noalign{\smallskip}  
 \multicolumn{5}{c}{\emph{Out-of-transit variability parameters}} \\  
 \noalign{\smallskip} \noalign{\smallskip}  
 Amplitude    &    $A_{\rm{NGTS}}$    &    \%         &    $\,\,\,\,\,0.00228\,^{+0.00033}_{-0.00027}$    &    $\,\,\,\,\,0.00228\,^{+0.00035}_{-0.00028}$    \\  [1.1ex]  
 Timescale    &    $l_{\rm{NGTS}}$    &    days         &    $\,\,\,\,\,2.76\,^{+0.29}_{-0.46}$    &    $\,\,\,\,\,2.76\,^{+0.29}_{-0.45}$    \\  [1.1ex]  
 White noise term    &    $\sigma_{\rm{NGTS}}$    &             &    $\,\,\,\,\,1.180 \pm 0.016$    &    $\,\,\,\,\,1.181 \pm 0.017$    \\  [1.1ex]  
 Amplitude    &    $A_{\rm{Euler}}$    &    \%         &    $\,\,\,\,\,0.035\,^{+0.056}_{-0.025}$    &    $\,\,\,\,\,0.065\,^{+0.179}_{-0.050}$    \\  [1.1ex]  
 Timescale    &    $l_{\rm{Euler}}$    &    days         &    $\,\,\,\,\,2.32\,^{+0.61}_{-0.83}$    &    $\,\,\,\,\,2.48\,^{+0.50}_{-0.82}$    \\  [1.1ex]  
 White noise term    &    $\sigma_{\rm{Euler}}$    &             &    $\,\,\,\,\,1.16 \pm 0.13$    &    $\,\,\,\,\,1.17 \pm 0.13$    \\  [1.1ex]  
 \noalign{\smallskip} \noalign{\smallskip} \noalign{\smallskip} \noalign{\smallskip} \noalign{\smallskip}  
 \multicolumn{5}{c}{\emph{Radial velocity parameters}} \\  
 \noalign{\smallskip} \noalign{\smallskip}  
 Systemic velocity    &    $V_{\rm{sys}}$    &    km\,s$^{-1}$        &    $\,\,\,\,\,97.1801\,^{+0.0071}_{-0.0078}$    &    $\,\,\,\,\,97.1809 \pm 0.0070$    \\  [1.1ex]  
 Primary RV semi-amplitude    &    $K_{\rm{pri}}$    &    km\,s$^{-1}$        &    $\,\,\,\,\,0.166 \pm 0.012$    &    $\,\,\,\,\,0.166 \pm 0.011$    \\  [1.1ex]  
 HARPS White noise term    &    $\sigma_{\rm{HARPS}}$    &    km\,s$^{-1}$        &    $\,\,\,\,\,0.0080\,^{+0.0116}_{-0.0057}$    &    $\,\,\,\,\,0.0071\,^{+0.0101}_{-0.0049}$    \\  [1.1ex]  
 \noalign{\smallskip} \noalign{\smallskip}\noalign{\smallskip}  
 \hline  
 \end{tabular}  
 \begin{list}{}{}  
 \item[* LDC = limb darkening coefficient]  
 \end{list} 
 \label{tab:gp}
 \end{table*} 

\begin{figure}
	\includegraphics[width=\columnwidth]{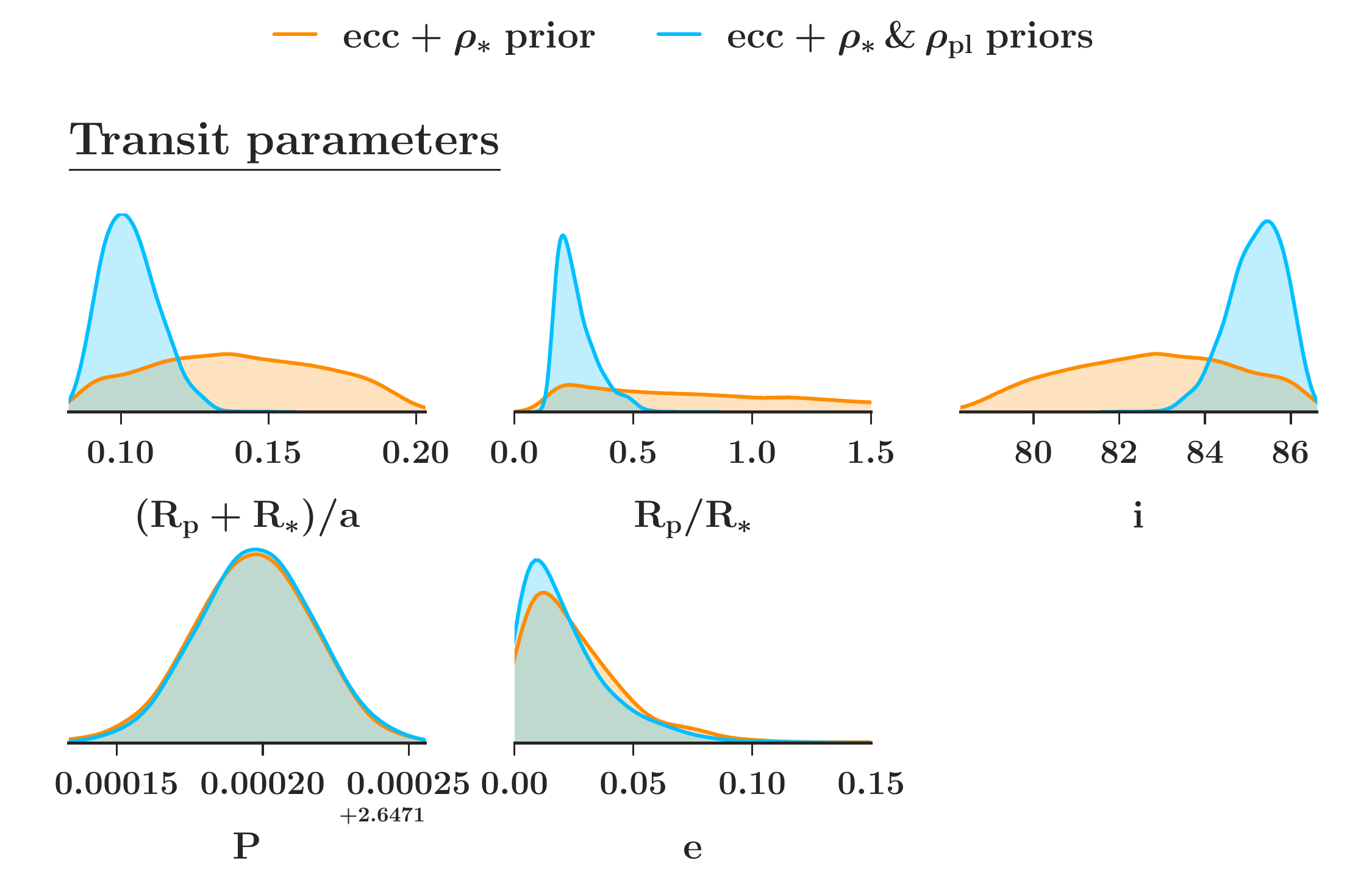}
   \caption{The posterior distributions for five key parameters from the \gpe\ global modelling with a prior on the stellar density, free-eccentricity and with/without (blue/orange) the adoption of the planet density prior.}
   \label{fig:posteriors}
\end{figure}

To illustrate the impact of this prior, Figure~\ref{fig:posteriors} shows the posterior distributions for five key model parameters with and without the planet density prior.  Most importantly we see the radius ratio ($\rpl/\rstar$) moves from an almost unconstrained posterior distributions to a more physically realistic distribution.  We caution that this approach means that the radius we measure for this planet is not as robust as for most transiting gas giants in the literature, and simply represents our best estimate based on all available data for this system and the population gas giant planets in general. 

Using these parameters from the \gpe\ global modelling and the stellar properties as derived in Section \ref{sub:stellar}, we derive the planet parameters for \Nplanet.  These planet parameters are set out in Table \ref{tab:planet}, and show that \Nplanet\ is a Jupiter mass (\Nmass\,\mjup) planet with a radius of \Nradius\,\rjup.

\begin{table}
	\centering
	\caption{Planetary Properties for \Nplanet}
	\begin{tabular}{lc} 
	Property	&	Value \\
	\hline
    P (days)		&	\Nperiod	\\
	T$_C$ (HJD)		&	\Ntc	\\
    T$_{14}$ (hours) & \Nduration\\
    $a/R_{*}$		& \Naoverr\\
    $b$ & \Nimpact    \\
	K (\ms) 	&\Nkamp	\\
    e 			& \Necc  	\\
    \mpl (\mjup)& \Nmass	\\
    \rpl (\rjup)& \Nradius  \\
    $\rho_{p}$ (\gccc) & \Ndensity\\
    a (AU) & \Nau \\
    T$_{eq}$ (K) & \NTeq	\\
  	Flux (\ergscm) & \Nflux	\\
	\hline
	\end{tabular}
    \label{tab:planet}
\end{table}


\subsection{Stellar Activity and Rotation}
\label{sub:activity}
Analysis of the H$\alpha$ line in the combined HARPS spectrum does not show any sign of emission that we would associate with activity, and there is no flaring detected in the \NGTS\ photometry.  The star is not coincident with any catalogued X-ray source, and there is no X-ray detection in the XMM-Slew survey \citep{saxton2008}. The XMM-Slew survey data provides a model-dependent upper limit of 5.7$\times 10^{-13}$~erg~s$^{-1}$~cm$^{-2}$ in the 0.2--2~keV band, and across the 0.2--12~keV band the upper limit is $1.4\times 10^{-12}$~erg~s$^{-1}$~cm$^{-2}$.  This is not surprising, as M-dwarfs of this effective temperature are expected to have flux levels below this upper limit \citep{Stelzer2013}. 

\begin{figure}
	\includegraphics[width=\columnwidth,angle=0]{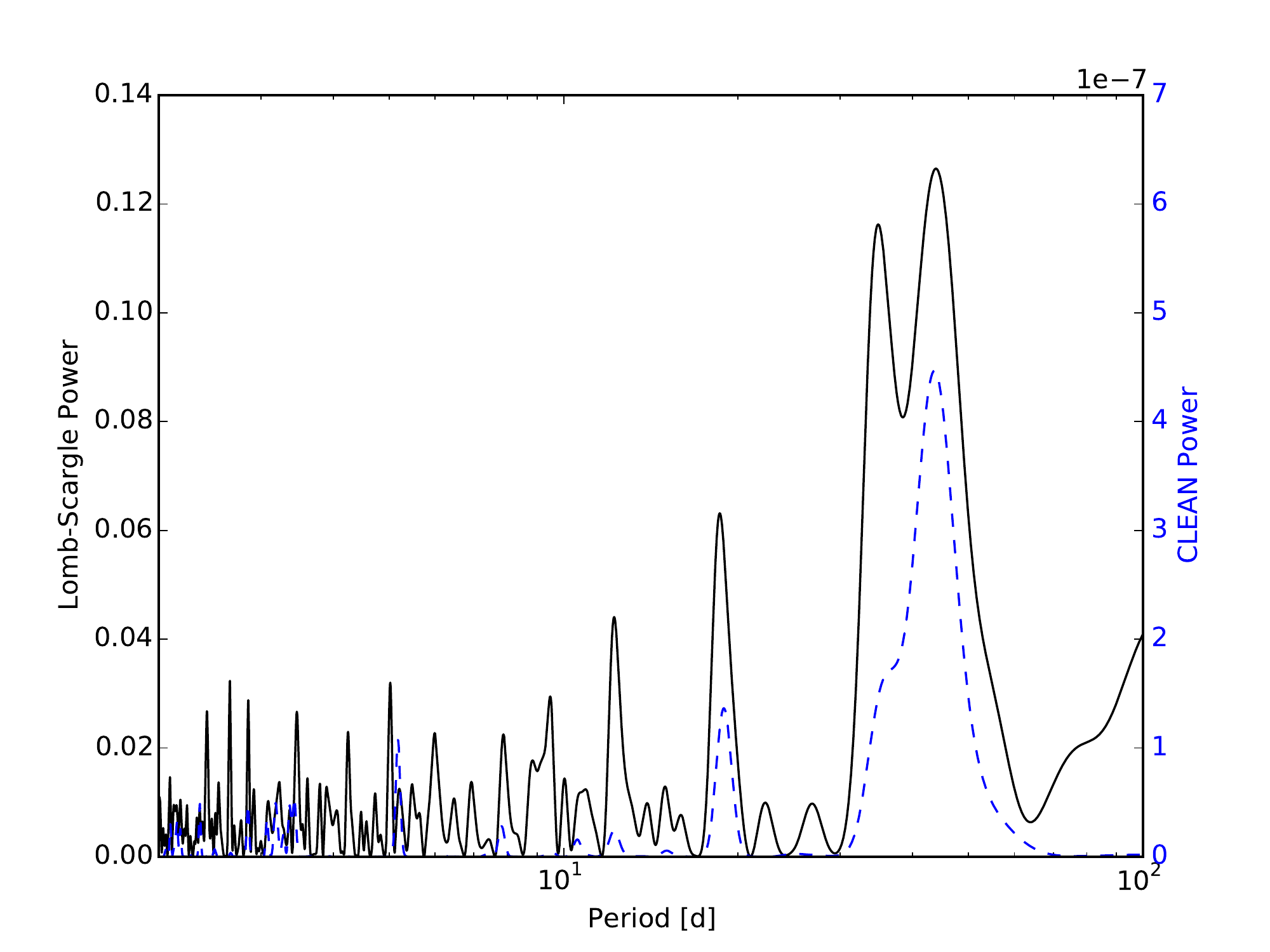}
    \caption{The Lombe-Scargle periodogram (black, continuous line) and the CLEAN periodogram (blue, dashed line) for \Nstar{} from the \NGTS{} photometric data.  The peak at \Nrotation\,d may be due to spot rotation.}
    \label{fig:rotation}
\end{figure}

M-dwarfs often have a larger fraction of spot coverage on their surface in comparison to FGK stars, and such spots may be revealed in photometric data as long term variability at the stellar rotation period.  To investigate this for \Nstar{} we ran the CLEAN algorithm \citep{Roberts1987} to identify periods and a Lomb-Scargle analysis \citep{lomb76} to estimate their significance using the \NGTS{} photometric data.  The result is set out in Figure~\ref{fig:rotation}. No significant signal was found. The broad peak at a period of approximately \Nrotation\,d corresponds to a signal with an amplitude of $\approx 1.5$ mmag, which is at the level of expected systematic noise. However, such a rotation period would be consistent with rotation rates for low-activity M-dwarfs \citep{mcquillan13}, and is similar to the other M-dwarf gas planet host, HATS-6, which has a probable rotation period of 35.1\,d \citep{hartman15}. 

\subsection{Archival Images}
\label{sub:archival}
Due to the relatively high proper motion of \Nstar{} (\NpropRA\,\masy, \NpropDec\,\masy), we are able to study archival images of \Nstar{} to help rule out blended background stars currently within the point-spread function of \Nstar.  The earliest image available is a photographic plate taken in 1976 December 26 as part of the SERC-J Survey on the UK Schmidt 1.2\,m telescope at Siding Spring Observatory, Australia.  The plate has been digitized as part of the STScI Digitized Sky Survey.  From this image we identify \Nstar{} at 05:30:51.543, -36:37:50.49.  Subsequent Digitized Sky Survey images are from 1985, 1992, and 1995.  2MASS imaged \Nstar{} in 1999.  Our own imaging is from 2016.  From these images we confirm the high proper motion, totalling 2.7\arcsec over 40 years.  This is consistent with the proper motion as reported in UCAC4 (see Table \ref{tab:stellar}).  The archival imaging allows us to rule out any background star at a significant flux level within the current \NGTS\ point-spread function.

\subsection{Kinematics}
\label{sub:kinematics}
Using the proper motion (R.A.:\NpropRA\,\masy, Dec.: \NpropDec\,\masy), the absolute radial velocity (\Ngamma\,\kms), and the estimated distance of \Ndist\,pc, we calculate the ($U_{LSR},V_{LSR},W_{LSR}$) galactic motion for \Nstar{} to be (\Ngalu, \Ngalv, \Ngalw)\,\kms.  This suggests that \Nstar\ is a member of the thick disk population, and may therefore be a very old system.  The major uncertainty in obtaining the galactic motion is the uncertainty in the distance to \Nstar.  It will be interesting to confirm the motion after the second GAIA data release \citep{GAIA}, which will provide a much more precise distance for \Nstar. 


\section{Discussion}
\label{sec:discussion}

\Nplanet{} is a Jupiter-mass planet orbiting an M0-dwarf star in a \Nperiod\,d period.  As such, it is just the third transiting gas giant to be discovered around an M-dwarf, and with $\mpl=\Nmass\,\mjup$
 it is easily the most massive planet known transiting an M-dwarf.  The two previously discovered examples are Kepler-45b \citep[\mpl=0.505\mjup, \rpl=0.96\rjup][]{johnson12} and HATS-6b \citep[\mpl=0.32\mjup, \rpl=1.00\rjup][]{hartman15}.  Due to the grazing nature of the transit, the radius of \Nplanet\ is not well constrained and is strongly influenced by the density prior we use.  Our hope is that 2-minute cadence data from the NASA {\em Transiting Exoplanet Survey Satellite} ({\em TESS}) mission \citep{ricker2014} will allow us to improve this estimate and model the system without the need for a prior on the planet density. 

\subsection{Giant Planet Formation around M-dwarfs}
\label{sub:formation}
Planet formation theory suggests that giant planets should be rarer around M-dwarfs than around FGK stars \citep{kennedy2008}. Dynamical time-scales around lower-mass stars are longer, slowing the process(es) of planet formation \citep[e.g.,][]{laughlin04}, and the mass budget is also lower \citep[as protoplanetary disc masses decrease approximately linearly with stellar mass;][]{Andrews13}.  Thus determining the frequency of giant planets around M-dwarfs should yield important new insights into the planet formation process.

It has long been known that for solar-type host stars the incidence of giant planets increases very sharply with increasing host star metallicity \citep{Fischer05}, and very few giant planets have been detected around stars with significantly sub-solar metallicity. However, the same correlation is not seen for Neptunes or super-Earths, whose incidence appears largely independent of metallicity \citep{Sousa08,buchhave15,jenkins17}. By robustly determining the metallicity of \Nstar{} and systems like this we will be able to probe if the correlation between metallicty and giant planet frequency remains true for low mass stars.

Probing the frequency of gas giants, and in particular, hot Jupiters orbiting M-dwarfs, is difficult.  The intrinsic low luminosity of M-dwarfs make them hard to monitor for radial velocity surveys.  The largest M-dwarf radial velocity survey, by \citet{bonfils13}, was able to put an upper limit of $\approx 1\%$ for P<10\,d and masses of $0.1<m$\,sin$i<1.0$\,\mjup.  However this result does not constrain M-dwarfs to have a lower frequency of hot Jupiters than FGK stars, which is robustly determined from the Kepler mission to be 0.43$\pm$0.05\% for P<10\,d and 0.55<\rpl<2 \citep{fressin13}.  Transit surveys face similar difficulties, with low luminosities meaning only a small population of M-dwarfs have been photometrically monitored for transits compared with FGK stars.  Small telescope surveys such as WASP or HATNet do not detect very faint or red objects and so do not have the ability to cover a significant population of M-dwarfs.  Surveys using larger telescopes but single-star monitoring, such as MEarth, do not monitor enough M-dwarfs to probe the frequency of hot Jupiters \citep{berta13}.  Kepler, with a 1\,m aperture, was able to monitor M-dwarfs in its field-of-view, but since it only covered 100 sq. deg., that population has been limited to approximately 4000 M-dwarfs \citep{dressing13}.  \NGTS\ is therefore currently the best placed survey to determine the frequency of hot Jupiters orbiting early M-dwarfs, monitoring approximately 20000 early-type M-dwarfs per year as part of the regular search campaigns.  However robust statistics will only be available after a larger number of fields (hence M-dwarfs) have been monitored over the coming years.

\subsection{\JWST\ Potential}
\label{sub:jwst}
The suite of instruments on-board the James Webb Space Telescope (\JWST) will allow for detailed characterisation of transiting exoplanets \citep{beichman14}.  In particular, transmission spectroscopy of short-period transiting gas giant planets is foreseen as one of the key tasks early in the mission lifetime \citep{stevenson16}.  

\begin{figure}
	\includegraphics[width=\columnwidth]{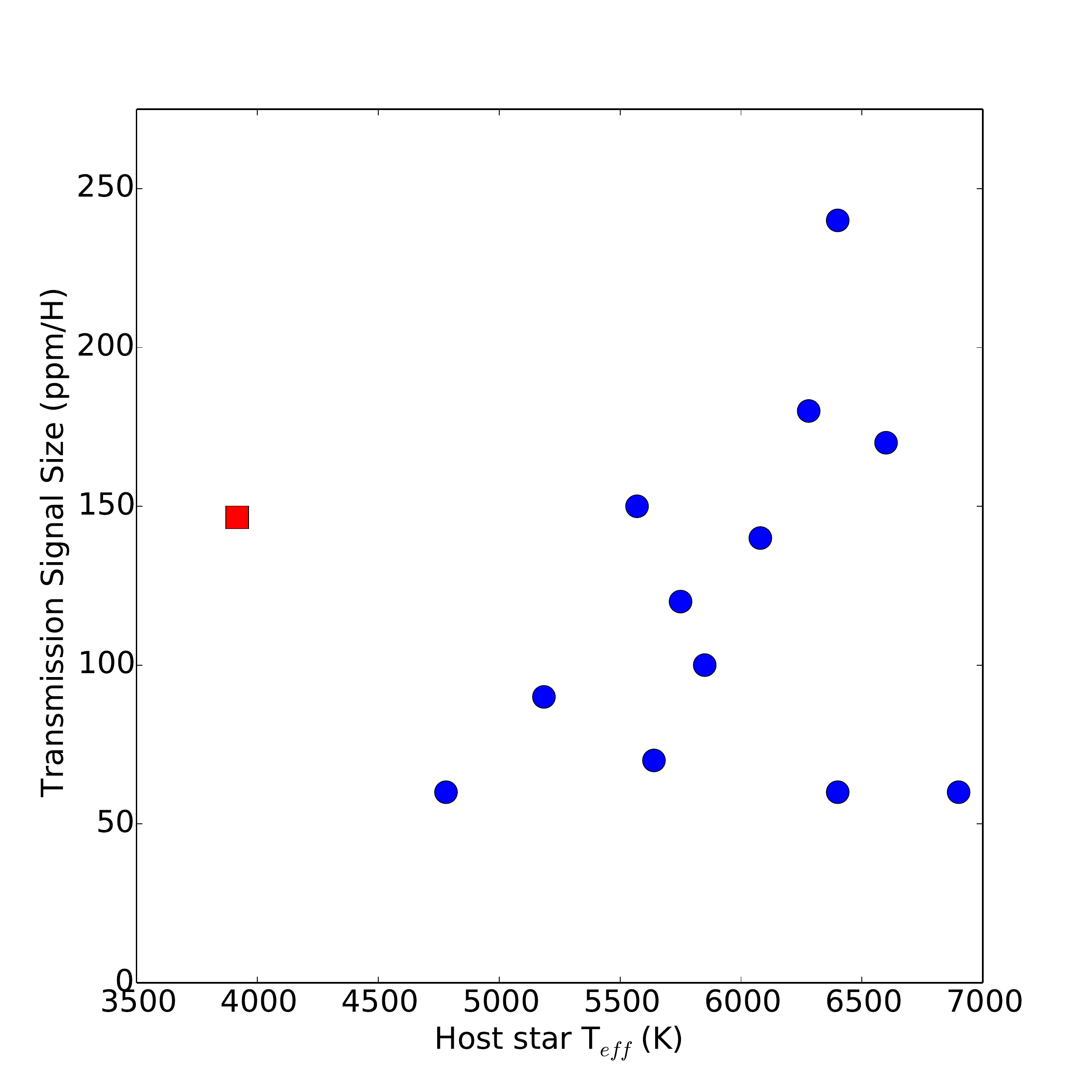}
    \caption{Signal size per scale height (in ppm/H) as a function of host star effective temperature for the \JWST\ community targets (blue circles; \citet{stevenson16}) and \Nplanet\ (red square).  The signal size for \Nplanet\ is comparable to that of the community targets, and it probes planet characterisation around a much cooler star than any of the community targets.  We note \Nstar\ is fainter by 2 magnitudes than the community targets.}
    \label{fig:jwst}
\end{figure}

At $J$=12.7, \Nstar{} is about 2 magnitudes fainter than the "community targets" proposed by \citet{stevenson16} for a \JWST\ Early Release Science Program.  However due to the large radius of the planet compared with its host star, the expected transmission signal is relatively high.  The signal size per scale height, $\Delta D$, assuming a cloud-free atmosphere at a constant equilibrium temperature, $T_{eq}$, is given by:
\begin{equation}
	\Delta D \approx \frac{2k_{B}T_{eq}R_{P}}{\mu g R_{S}^{2}}
\end{equation}

\noindent where $R_{P}$ and $R_{S}$ are the planetary and stellar radii, respectively, $\mu$ is the mean molecular weight, $g$ is the planet's gravity, and $k_{B}$ is the Boltzmann constant.  Assuming a mean molecular weight of of 2.2u, and applying a factor of 0.5 to account for the transit being grazing, we calculate a signal size of $\Delta D=$\Nsignalsize\ for \Nplanet.  We see \Nplanet\ has a signal comparable to the \JWST\ community targets, though with a very different host star mass (see Figure~\ref{fig:jwst}).  In addition, the short period of \Nplanet\ (P=\Nperiod d) means there will be many available transits per year, and its location on the sky (\NRA, \NDec) means that it lies $\approx$30\,deg from the \JWST\ southern continuous viewing zone, which will allow for over 200\,d per year of observing opportunity from \JWST.  The complicating factor here is the fact that \Nplanet\ is undergoing a grazing transit, which means that robust atmospheric transmission spectroscopy will require a better understanding of the limb darkening parameters for this host star than is typically the case for a more centrally crossing transit.  It also means the transit is shorter in duration, so there is less in-transit time per orbit for transmission spectroscopy.

\subsection{Possible Thick Disk Membership}
\label{sub:age}
As discussed in Section \ref{sub:kinematics}, the kinematics of \Nstar{} is consistent with a star from the thick disk population.  Additionally this M-dwarf appears to be slowly rotating and shows no signs of activity from both photometric and spectroscopic indicators.  Therefore it seems likely that this system is very old, indicating that giant planet formation around M-dwarfs was possible from an early point in the Galaxy's history.  However IR spectroscopy will help determine if the stellar metallicity confirms this picture, as will the more robust distance measurement which will be part of GAIA DR2.

\section{Conclusions}
The discovery of \Nplanet\ demonstrates the capability of \NGTS\ to probe early M-dwarfs for transiting planets - the result of the 20cm aperture telescopes coupled to red-sensitive CCDs.  In the full course of the survey, enough early M-dwarfs will be monitored to allow us to provide statistics for these host stars such as the frequency of hot Jupiters around early M-dwarfs. 

The {\em TESS} mission \citep{ricker2014} plans to focus on early M-dwarfs in order to be able to detect potentially habitable planets.  \Nstar\ has an $i$-band magnitude of $14.38 \pm 0.02$, which for {\em TESS} will be in a sky- and readout-limited regime and therefore will only have an expected precision of 3mmag per hour (c.f. the \NGTS\ light curve is much \textit{better} than this expected \TESS\ precision at this magnitude).  However given the depth of the \Nplanet\ transit and the approximately continuous observing window function from {\em TESS}, planets similar to \Nplanet\ should be picked up very easily by \TESS\, promising to shed more light on giant planets around low-mass stars.

\section*{Acknowledgements}
Based on data collected under the NGTS project at the ESO La Silla Paranal Observatory.  The NGTS facility is operated by the consortium institutes with support from the UK Science and Technology Facilities Council (STFC)  project ST/M001962/1. 

Contributions at the University of Geneva by DB, FB, BC, LM, and SU were carried out within the framework of the National Centre for Competence in Research "PlanetS" supported by the Swiss National Science Foundation (SNSF).
The contributions at the University of Warwick by PJW, RGW, DLP, FF, DA, BTG and TL have been supported by STFC through consolidated grants ST/L000733/1 and ST/P000495/1. 
The contributions at the University of Leicester by MGW and MRB have been supported by STFC through consolidated grant ST/N000757/1.

CAW acknowledges support from the STFC grant ST/P000312/1.
TL was also supported by STFC studentship 1226157.
MNG is supported by the STFC award reference 1490409 as well as the Isaac Newton Studentship.
JSJ acknowledges support by Fondecyt grant 1161218 and partial support by CATA-Basal (PB06, CONICYT).
This project has received funding from the European Research Council (ERC) under the European Union's Horizon 2020 research and innovation programme (grant agreement No 681601).
The research leading to these results has received funding from the European Research Council under the European Union's Seventh Framework Programme (FP/2007-2013) / ERC Grant Agreement n. 320964 (WDTracer).
We thank Andy Read (University of Leicester) for useful discussions on the XMM-Slew Survey data. 




	\bibliographystyle{mnras}
\bibliography{paper} 

\begin{thebibliography}{}
\makeatletter
\relax
\def\mn@urlcharsother{\let\do\@makeother \do\$\do\&\do\#\do\^\do\_\do\%\do\~}
\def\mn@doi{\begingroup\mn@urlcharsother \@ifnextchar [ {\mn@doi@}
  {\mn@doi@[]}}
\def\mn@doi@[#1]#2{\def\@tempa{#1}\ifx\@tempa\@empty \href
  {http://dx.doi.org/#2} {doi:#2}\else \href {http://dx.doi.org/#2} {#1}\fi
  \endgroup}
\def\mn@eprint#1#2{\mn@eprint@#1:#2::\@nil}
\def\mn@eprint@arXiv#1{\href {http://arxiv.org/abs/#1} {{\tt arXiv:#1}}}
\def\mn@eprint@dblp#1{\href {http://dblp.uni-trier.de/rec/bibtex/#1.xml}
  {dblp:#1}}
\def\mn@eprint@#1:#2:#3:#4\@nil{\def\@tempa {#1}\def\@tempb {#2}\def\@tempc
  {#3}\ifx \@tempc \@empty \let \@tempc \@tempb \let \@tempb \@tempa \fi \ifx
  \@tempb \@empty \def\@tempb {arXiv}\fi \@ifundefined
  {mn@eprint@\@tempb}{\@tempb:\@tempc}{\expandafter \expandafter \csname
  mn@eprint@\@tempb\endcsname \expandafter{\@tempc}}}

\bibitem[\protect\citeauthoryear{{Akeson} et~al.,}{{Akeson}
  et~al.}{2013}]{akeson13}
{Akeson} R.~L.,  et~al., 2013, \mn@doi [\pasp] {10.1086/672273}, \href
  {http://adsabs.harvard.edu/abs/2013PASP..125..989A} {125, 989}

\bibitem[\protect\citeauthoryear{{Allard}, {Homeier}  \& {Freytag}}{{Allard}
  et~al.}{2011}]{Allard2011}
{Allard} F.,  {Homeier} D.,   {Freytag} B.,  2011, in {Johns-Krull} C.,
  {Browning} M.~K.,   {West} A.~A.,  eds,  Astronomical Society of the Pacific
  Conference Series Vol. 448, 16th Cambridge Workshop on Cool Stars, Stellar
  Systems, and the Sun. p.~91 (\mn@eprint {arXiv} {1011.5405})

\bibitem[\protect\citeauthoryear{{Andrews}, {Rosenfeld}, {Kraus}  \&
  {Wilner}}{{Andrews} et~al.}{2013}]{Andrews13}
{Andrews} S.~M.,  {Rosenfeld} K.~A.,  {Kraus} A.~L.,   {Wilner} D.~J.,  2013,
  \mn@doi [\apj] {10.1088/0004-637X/771/2/129}, \href
  {http://adsabs.harvard.edu/abs/2013ApJ...771..129A} {771, 129}

\bibitem[\protect\citeauthoryear{{Anglada-Escud{\'e}}
  et~al.,}{{Anglada-Escud{\'e}} et~al.}{2016}]{2016Natur.536..437A}
{Anglada-Escud{\'e}} G.,  et~al., 2016, \mn@doi [\nat] {10.1038/nature19106},
  \href {http://adsabs.harvard.edu/abs/2016Natur.536..437A} {536, 437}

\bibitem[\protect\citeauthoryear{{Astudillo-Defru} et~al.,}{{Astudillo-Defru}
  et~al.}{2017}]{Astudillo2017}
{Astudillo-Defru} N.,  et~al., 2017, \mn@doi [\aap]
  {10.1051/0004-6361/201630153}, \href
  {http://adsabs.harvard.edu/abs/2017A%26A...602A..88A} {602, A88}

\bibitem[\protect\citeauthoryear{{Beichman} et~al.,}{{Beichman}
  et~al.}{2014}]{beichman14}
{Beichman} C.,  et~al., 2014, \mn@doi [\pasp] {10.1086/679566}, \href
  {http://adsabs.harvard.edu/abs/2014PASP..126.1134B} {126, 1134}

\bibitem[\protect\citeauthoryear{{Benedict} et~al.,}{{Benedict}
  et~al.}{2016}]{Benedict16}
{Benedict} G.~F.,  et~al., 2016, \mn@doi [\aj] {10.3847/0004-6256/152/5/141},
  \href {http://adsabs.harvard.edu/abs/2016AJ....152..141B} {152, 141}

\bibitem[\protect\citeauthoryear{{Berdi{\~n}as}}{{Berdi{\~n}as}}{2016}]{berdinas2016}
{Berdi{\~n}as} Z.~M.,  2016, PhD thesis, Instituto de Astrof{\'{\i}}sica de
  Andaluc{\'{\i}}a-CSIC; Universidad de Granada, \mn@doi{10.5281/zenodo.438330}

\bibitem[\protect\citeauthoryear{{Berta-Thompson} et~al.,}{{Berta-Thompson}
  et~al.}{2015}]{BertaThompson2015}
{Berta-Thompson} Z.~K.,  et~al., 2015, \mn@doi [\nat] {10.1038/nature15762},
  \href {http://adsabs.harvard.edu/abs/2015Natur.527..204B} {527, 204}

\bibitem[\protect\citeauthoryear{{Berta}, {Irwin}  \& {Charbonneau}}{{Berta}
  et~al.}{2013}]{berta13}
{Berta} Z.~K.,  {Irwin} J.,   {Charbonneau} D.,  2013, \mn@doi [\apj]
  {10.1088/0004-637X/775/2/91}, \href
  {http://adsabs.harvard.edu/abs/2013ApJ...775...91B} {775, 91}

\bibitem[\protect\citeauthoryear{{Bertin} \& {Arnouts}}{{Bertin} \&
  {Arnouts}}{1996}]{1996A&AS..117..393B}
{Bertin} E.,  {Arnouts} S.,  1996, \mn@doi [\aaps] {10.1051/aas:1996164}, \href
  {http://adsabs.harvard.edu/abs/1996A%26AS..117..393B} {117, 393}

\bibitem[\protect\citeauthoryear{{Bonfils} et~al.,}{{Bonfils}
  et~al.}{2005}]{Bonfils2005}
{Bonfils} X.,  et~al., 2005, \mn@doi [\aap] {10.1051/0004-6361:200500193},
  \href {http://adsabs.harvard.edu/abs/2005A%26A...443L..15B} {443, L15}

\bibitem[\protect\citeauthoryear{{Bonfils} et~al.,}{{Bonfils}
  et~al.}{2013}]{bonfils13}
{Bonfils} X.,  et~al., 2013, \mn@doi [\aap] {10.1051/0004-6361/201014704},
  \href {http://adsabs.harvard.edu/abs/2013A%26A...549A.109B} {549, A109}

\bibitem[\protect\citeauthoryear{{Buchhave} \& {Latham}}{{Buchhave} \&
  {Latham}}{2015}]{buchhave15}
{Buchhave} L.~A.,  {Latham} D.~W.,  2015, \mn@doi [\apj]
  {10.1088/0004-637X/808/2/187}, \href
  {http://adsabs.harvard.edu/abs/2015ApJ...808..187B} {808, 187}

\bibitem[\protect\citeauthoryear{{Charbonneau} et~al.,}{{Charbonneau}
  et~al.}{2009}]{Charbonneau2009}
{Charbonneau} D.,  et~al., 2009, \mn@doi [\nat] {10.1038/nature08679}, \href
  {http://adsabs.harvard.edu/abs/2009Natur.462..891C} {462, 891}

\bibitem[\protect\citeauthoryear{{Chazelas} et~al.,}{{Chazelas}
  et~al.}{2012}]{Chazelas2012}
{Chazelas} B.,  et~al., 2012, in Ground-based and Airborne Telescopes IV. p.
  84440E, \mn@doi{10.1117/12.925755}

\bibitem[\protect\citeauthoryear{{David} et~al.,}{{David}
  et~al.}{2016}]{David2016}
{David} T.~J.,  et~al., 2016, \mn@doi [\nat] {10.1038/nature18293}, \href
  {http://adsabs.harvard.edu/abs/2016Natur.534..658D} {534, 658}

\bibitem[\protect\citeauthoryear{{Demory} \& {Seager}}{{Demory} \&
  {Seager}}{2011}]{demory11}
{Demory} B.-O.,  {Seager} S.,  2011, \mn@doi [\apjs]
  {10.1088/0067-0049/197/1/12}, \href
  {http://adsabs.harvard.edu/abs/2011ApJS..197...12D} {197, 12}

\bibitem[\protect\citeauthoryear{{Dressing} \& {Charbonneau}}{{Dressing} \&
  {Charbonneau}}{2013}]{dressing13}
{Dressing} C.~D.,  {Charbonneau} D.,  2013, \mn@doi [\apj]
  {10.1088/0004-637X/767/1/95}, \href
  {http://adsabs.harvard.edu/abs/2013ApJ...767...95D} {767, 95}

\bibitem[\protect\citeauthoryear{{Fischer} \& {Valenti}}{{Fischer} \&
  {Valenti}}{2005}]{Fischer05}
{Fischer} D.~A.,  {Valenti} J.,  2005, \mn@doi [\apj] {10.1086/428383}, \href
  {http://adsabs.harvard.edu/abs/2005ApJ...622.1102F} {622, 1102}

\bibitem[\protect\citeauthoryear{{Fressin} et~al.,}{{Fressin}
  et~al.}{2013}]{fressin13}
{Fressin} F.,  et~al., 2013, \mn@doi [\apj] {10.1088/0004-637X/766/2/81}, \href
  {http://adsabs.harvard.edu/abs/2013ApJ...766...81F} {766, 81}

\bibitem[\protect\citeauthoryear{{Gaia Collaboration} et~al.,}{{Gaia
  Collaboration} et~al.}{2016}]{GAIA}
{Gaia Collaboration} et~al., 2016, \mn@doi [\aap]
  {10.1051/0004-6361/201629512}, \href
  {http://adsabs.harvard.edu/abs/2016A%26A...595A...2G} {595, A2}

\bibitem[\protect\citeauthoryear{{Gillen}, {Hillenbrand}, {David}, {Aigrain},
  {Rebull}, {Stauffer}, {Cody}  \& {Queloz}}{{Gillen} et~al.}{2017}]{Gillen17}
{Gillen} E.,  {Hillenbrand} L.~A.,  {David} T.~J.,  {Aigrain} S.,  {Rebull} L.,
   {Stauffer} J.,  {Cody} A.~M.,   {Queloz} D.,  2017, \mn@doi [\apj]
  {10.3847/1538-4357/849/1/11}, \href
  {http://adsabs.harvard.edu/abs/2017arXiv170603084G} {849, 11}

\bibitem[\protect\citeauthoryear{{Gillon} et~al.,}{{Gillon}
  et~al.}{2017}]{2017Natur.542..456G}
{Gillon} M.,  et~al., 2017, \mn@doi [\nat] {10.1038/nature21360}, \href
  {http://adsabs.harvard.edu/abs/2017Natur.542..456G} {542, 456}

\bibitem[\protect\citeauthoryear{{G{\"u}nther}, {Queloz}, {Demory}  \&
  {Bouchy}}{{G{\"u}nther} et~al.}{2017a}]{Guenther17a}
{G{\"u}nther} M.~N.,  {Queloz} D.,  {Demory} B.-O.,   {Bouchy} F.,  2017a,
  \mn@doi [\mnras] {10.1093/mnras/stw2908}, \href
  {http://adsabs.harvard.edu/abs/2017MNRAS.465.3379G} {465, 3379}

\bibitem[\protect\citeauthoryear{{G{\"u}nther} et~al.,}{{G{\"u}nther}
  et~al.}{2017b}]{Guenther17b}
{G{\"u}nther} M.~N.,  et~al., 2017b, \mn@doi [\mnras] {10.1093/mnras/stx1920},
  472, 295

\bibitem[\protect\citeauthoryear{{Hartman} et~al.,}{{Hartman}
  et~al.}{2015}]{hartman15}
{Hartman} J.~D.,  et~al., 2015, \mn@doi [\aj] {10.1088/0004-6256/149/5/166},
  \href {http://adsabs.harvard.edu/abs/2015AJ....149..166H} {149, 166}

\bibitem[\protect\citeauthoryear{{Hatzes} \& {Rauer}}{{Hatzes} \&
  {Rauer}}{2015}]{hatzes15}
{Hatzes} A.~P.,  {Rauer} H.,  2015, \mn@doi [\apjl]
  {10.1088/2041-8205/810/2/L25}, \href
  {http://adsabs.harvard.edu/abs/2015ApJ...810L..25H} {810, L25}

\bibitem[\protect\citeauthoryear{{Henden} \& {Munari}}{{Henden} \&
  {Munari}}{2014}]{APASS}
{Henden} A.,  {Munari} U.,  2014, Contributions of the Astronomical Observatory
  Skalnate Pleso, \href {http://adsabs.harvard.edu/abs/2014CoSka..43..518H}
  {43, 518}

\bibitem[\protect\citeauthoryear{{Henry}, {Jao}, {Subasavage}, {Beaulieu},
  {Ianna}, {Costa}  \& {M{\'e}ndez}}{{Henry} et~al.}{2006}]{henry06}
{Henry} T.~J.,  {Jao} W.-C.,  {Subasavage} J.~P.,  {Beaulieu} T.~D.,  {Ianna}
  P.~A.,  {Costa} E.,   {M{\'e}ndez} R.~A.,  2006, \mn@doi [\aj]
  {10.1086/508233}, \href {http://adsabs.harvard.edu/abs/2006AJ....132.2360H}
  {132, 2360}

\bibitem[\protect\citeauthoryear{{Jenkins} et~al.,}{{Jenkins}
  et~al.}{2017}]{jenkins17}
{Jenkins} J.~S.,  et~al., 2017, \mn@doi [\mnras] {10.1093/mnras/stw2811}, \href
  {http://adsabs.harvard.edu/abs/2017MNRAS.466..443J} {466, 443}

\bibitem[\protect\citeauthoryear{{Johnson} et~al.,}{{Johnson}
  et~al.}{2012}]{johnson12}
{Johnson} J.~A.,  et~al., 2012, \mn@doi [\aj] {10.1088/0004-6256/143/5/111},
  \href {http://adsabs.harvard.edu/abs/2012AJ....143..111J} {143, 111}

\bibitem[\protect\citeauthoryear{{Kennedy} \& {Kenyon}}{{Kennedy} \&
  {Kenyon}}{2008}]{kennedy2008}
{Kennedy} G.~M.,  {Kenyon} S.~J.,  2008, \mn@doi [\apj] {10.1086/524130}, \href
  {http://adsabs.harvard.edu/abs/2008ApJ...673..502K} {673, 502}

\bibitem[\protect\citeauthoryear{{Kov{\'a}cs}, {Zucker}  \&
  {Mazeh}}{{Kov{\'a}cs} et~al.}{2002}]{Kovacs2002}
{Kov{\'a}cs} G.,  {Zucker} S.,   {Mazeh} T.,  2002, \mn@doi [\aap]
  {10.1051/0004-6361:20020802}, \href
  {http://adsabs.harvard.edu/abs/2002A%26A...391..369K} {391, 369}

\bibitem[\protect\citeauthoryear{{Laughlin}, {Bodenheimer}  \&
  {Adams}}{{Laughlin} et~al.}{2004}]{laughlin04}
{Laughlin} G.,  {Bodenheimer} P.,   {Adams} F.~C.,  2004, \mn@doi [\apjl]
  {10.1086/424384}, \href {http://adsabs.harvard.edu/abs/2004ApJ...612L..73L}
  {612, L73}

\bibitem[\protect\citeauthoryear{{L{\'e}pine}, {Hilton}, {Mann}, {Wilde},
  {Rojas-Ayala}, {Cruz}  \& {Gaidos}}{{L{\'e}pine} et~al.}{2013}]{lepine13}
{L{\'e}pine} S.,  {Hilton} E.~J.,  {Mann} A.~W.,  {Wilde} M.,  {Rojas-Ayala}
  B.,  {Cruz} K.~L.,   {Gaidos} E.,  2013, \mn@doi [\aj]
  {10.1088/0004-6256/145/4/102}, \href
  {http://adsabs.harvard.edu/abs/2013AJ....145..102L} {145, 102}

\bibitem[\protect\citeauthoryear{{Lomb}}{{Lomb}}{1976}]{lomb76}
{Lomb} N.~R.,  1976, \mn@doi [\apss] {10.1007/BF00648343}, \href
  {http://adsabs.harvard.edu/abs/1976Ap%26SS..39..447L} {39, 447}

\bibitem[\protect\citeauthoryear{{Mann}, {Feiden}, {Gaidos}, {Boyajian}  \&
  {von Braun}}{{Mann} et~al.}{2015}]{Mann15}
{Mann} A.~W.,  {Feiden} G.~A.,  {Gaidos} E.,  {Boyajian} T.,   {von Braun} K.,
  2015, \mn@doi [\apj] {10.1088/0004-637X/804/1/64}, \href
  {http://adsabs.harvard.edu/abs/2015ApJ...804...64M} {804, 64}

\bibitem[\protect\citeauthoryear{{Mayor} et~al.,}{{Mayor}
  et~al.}{2003a}]{Mayor2003}
{Mayor} M.,  et~al., 2003a, The Messenger, \href
  {http://adsabs.harvard.edu/abs/2003Msngr.114...20M} {114, 20}

\bibitem[\protect\citeauthoryear{{Mayor} et~al.,}{{Mayor}
  et~al.}{2003b}]{2003Msngr.114...20M}
{Mayor} M.,  et~al., 2003b, The Messenger, \href
  {http://adsabs.harvard.edu/abs/2003Msngr.114...20M} {114, 20}

\bibitem[\protect\citeauthoryear{{McCormac} et~al.,}{{McCormac}
  et~al.}{2017}]{McCormac2017}
{McCormac} J.,  et~al., 2017, \mn@doi [\pasp]
  {10.1088/1538-3873/129/972/025002}, \href
  {http://adsabs.harvard.edu/abs/2017PASP..129b5002M} {129, 025002}

\bibitem[\protect\citeauthoryear{{McQuillan}, {Aigrain}  \&
  {Mazeh}}{{McQuillan} et~al.}{2013}]{mcquillan13}
{McQuillan} A.,  {Aigrain} S.,   {Mazeh} T.,  2013, \mn@doi [\mnras]
  {10.1093/mnras/stt536}, \href
  {http://adsabs.harvard.edu/abs/2013MNRAS.432.1203M} {432, 1203}

\bibitem[\protect\citeauthoryear{{Meyer}, {Amara}, {Reggiani}  \&
  {Quanz}}{{Meyer} et~al.}{2017}]{meyer17}
{Meyer} M.~R.,  {Amara} A.,  {Reggiani} M.,   {Quanz} S.~P.,  2017, preprint,
  \href {http://adsabs.harvard.edu/abs/2017arXiv170705256M} {} (\mn@eprint
  {arXiv} {1707.05256})

\bibitem[\protect\citeauthoryear{{Nutzman} \& {Charbonneau}}{{Nutzman} \&
  {Charbonneau}}{2008}]{Nutzman2008}
{Nutzman} P.,  {Charbonneau} D.,  2008, \mn@doi [\pasp] {10.1086/533420}, \href
  {http://adsabs.harvard.edu/abs/2008PASP..120..317N} {120, 317}

\bibitem[\protect\citeauthoryear{{Pepper} et~al.,}{{Pepper}
  et~al.}{2017}]{pepper17}
{Pepper} J.,  et~al., 2017, \mn@doi [\aj] {10.3847/1538-3881/aa62ab}, \href
  {http://adsabs.harvard.edu/abs/2017AJ....153..177P} {153, 177}

\bibitem[\protect\citeauthoryear{{Rebassa-Mansergas}, {G{\"a}nsicke},
  {Rodr{\'{\i}}guez-Gil}, {Schreiber}  \& {Koester}}{{Rebassa-Mansergas}
  et~al.}{2007}]{rebassa-mansergasetal07-1}
{Rebassa-Mansergas} A.,  {G{\"a}nsicke} B.~T.,  {Rodr{\'{\i}}guez-Gil} P.,
  {Schreiber} M.~R.,   {Koester} D.,  2007, \mn@doi [\mnras]
  {10.1111/j.1365-2966.2007.12288.x}, \href {2007MNRAS.382.1377R} {382, 1377}

\bibitem[\protect\citeauthoryear{{Ricker} et~al.,}{{Ricker}
  et~al.}{2014}]{ricker2014}
{Ricker} G.~R.,  et~al., 2014, in Space Telescopes and Instrumentation 2014:
  Optical, Infrared, and Millimeter Wave. p. 914320 (\mn@eprint {arXiv}
  {1406.0151}), \mn@doi{10.1117/12.2063489}

\bibitem[\protect\citeauthoryear{{Roberts}, {Lehar}  \& {Dreher}}{{Roberts}
  et~al.}{1987}]{Roberts1987}
{Roberts} D.~H.,  {Lehar} J.,   {Dreher} J.~W.,  1987, \mn@doi [\aj]
  {10.1086/114383}, \href {http://adsabs.harvard.edu/abs/1987AJ.....93..968R}
  {93, 968}

\bibitem[\protect\citeauthoryear{{Robin}, {Reyl{\'e}}, {Derri{\`e}re}  \&
  {Picaud}}{{Robin} et~al.}{2003}]{robin03}
{Robin} A.~C.,  {Reyl{\'e}} C.,  {Derri{\`e}re} S.,   {Picaud} S.,  2003,
  \mn@doi [\aap] {10.1051/0004-6361:20031117}, \href
  {http://adsabs.harvard.edu/abs/2003A%26A...409..523R} {409, 523}

\bibitem[\protect\citeauthoryear{{Rojas-Ayala}, {Covey}, {Muirhead}  \&
  {Lloyd}}{{Rojas-Ayala} et~al.}{2010}]{rojas10}
{Rojas-Ayala} B.,  {Covey} K.~R.,  {Muirhead} P.~S.,   {Lloyd} J.~P.,  2010,
  \mn@doi [\apjl] {10.1088/2041-8205/720/1/L113}, \href
  {http://adsabs.harvard.edu/abs/2010ApJ...720L.113R} {720, L113}

\bibitem[\protect\citeauthoryear{{Rowe} et~al.,}{{Rowe}
  et~al.}{2014}]{Rowe2014}
{Rowe} J.~F.,  et~al., 2014, \mn@doi [\apj] {10.1088/0004-637X/784/1/45}, \href
  {http://adsabs.harvard.edu/abs/2014ApJ...784...45R} {784, 45}

\bibitem[\protect\citeauthoryear{{Savcheva}, {West}  \& {Bochanski}}{{Savcheva}
  et~al.}{2014}]{savcheva2014}
{Savcheva} A.~S.,  {West} A.~A.,   {Bochanski} J.~J.,  2014, \mn@doi [\apj]
  {10.1088/0004-637X/794/2/145}, \href
  {http://adsabs.harvard.edu/abs/2014ApJ...794..145S} {794, 145}

\bibitem[\protect\citeauthoryear{{Saxton}, {Read}, {Esquej}, {Freyberg},
  {Altieri}  \& {Bermejo}}{{Saxton} et~al.}{2008}]{saxton2008}
{Saxton} R.~D.,  {Read} A.~M.,  {Esquej} P.,  {Freyberg} M.~J.,  {Altieri} B.,
   {Bermejo} D.,  2008, \mn@doi [\aap] {10.1051/0004-6361:20079193}, \href
  {http://adsabs.harvard.edu/abs/2008A%26A...480..611S} {480, 611}

\bibitem[\protect\citeauthoryear{{Skrutskie} et~al.,}{{Skrutskie}
  et~al.}{2006}]{2MASS}
{Skrutskie} M.~F.,  et~al., 2006, \mn@doi [\aj] {10.1086/498708}, \href
  {http://adsabs.harvard.edu/abs/2006AJ....131.1163S} {131, 1163}

\bibitem[\protect\citeauthoryear{{Sousa} et~al.,}{{Sousa}
  et~al.}{2008}]{Sousa08}
{Sousa} S.~G.,  et~al., 2008, \mn@doi [\aap] {10.1051/0004-6361:200809698},
  \href {http://adsabs.harvard.edu/abs/2008A%26A...487..373S} {487, 373}

\bibitem[\protect\citeauthoryear{{Stelzer}, {Marino}, {Micela},
  {L{\'o}pez-Santiago}  \& {Liefke}}{{Stelzer} et~al.}{2013}]{Stelzer2013}
{Stelzer} B.,  {Marino} A.,  {Micela} G.,  {L{\'o}pez-Santiago} J.,   {Liefke}
  C.,  2013, \mn@doi [\mnras] {10.1093/mnras/stt225}, \href
  {http://adsabs.harvard.edu/abs/2013MNRAS.431.2063S} {431, 2063}

\bibitem[\protect\citeauthoryear{{Stevenson} et~al.,}{{Stevenson}
  et~al.}{2016}]{stevenson16}
{Stevenson} K.~B.,  et~al., 2016, \mn@doi [\pasp]
  {10.1088/1538-3873/128/967/094401}, \href
  {http://adsabs.harvard.edu/abs/2016PASP..128i4401S} {128, 094401}

\bibitem[\protect\citeauthoryear{{Tamuz}, {Mazeh}  \& {Zucker}}{{Tamuz}
  et~al.}{2005}]{Tamuz2005}
{Tamuz} O.,  {Mazeh} T.,   {Zucker} S.,  2005, \mn@doi [\mnras]
  {10.1111/j.1365-2966.2004.08585.x}, \href
  {http://adsabs.harvard.edu/abs/2005MNRAS.356.1466T} {356, 1466}

\bibitem[\protect\citeauthoryear{{Triaud} et~al.,}{{Triaud}
  et~al.}{2013}]{wasp80}
{Triaud} A.~H.~M.~J.,  et~al., 2013, \mn@doi [\aap]
  {10.1051/0004-6361/201220900}, \href
  {http://adsabs.harvard.edu/abs/2013A%26A...551A..80T} {551, A80}

\bibitem[\protect\citeauthoryear{{Wheatley} et~al.,}{{Wheatley}
  et~al.}{2013}]{2013EPJWC..4713002W}
{Wheatley} P.~J.,  et~al., 2013, in European Physical Journal Web of
  Conferences. p. 13002 (\mn@eprint {arXiv} {1302.6592}),
  \mn@doi{10.1051/epjconf/20134713002}

\bibitem[\protect\citeauthoryear{{Wheatley} et~al.,}{{Wheatley}
  et~al.}{2017}]{project2017}
{Wheatley} P.,  et~al., 2017, \mnras, \href
  {http://adsabs.harvard.edu/abs/2017MNRAS....999.999W} {accepted}

\bibitem[\protect\citeauthoryear{{Wright} et~al.,}{{Wright}
  et~al.}{2010}]{WISE}
{Wright} E.~L.,  et~al., 2010, \mn@doi [\aj] {10.1088/0004-6256/140/6/1868},
  \href {http://adsabs.harvard.edu/abs/2010AJ....140.1868W} {140, 1868}

\bibitem[\protect\citeauthoryear{{Zacharias}, {Finch}, {Girard}, {Henden},
  {Bartlett}, {Monet}  \& {Zacharias}}{{Zacharias} et~al.}{2013}]{UCAC}
{Zacharias} N.,  {Finch} C.~T.,  {Girard} T.~M.,  {Henden} A.,  {Bartlett}
  J.~L.,  {Monet} D.~G.,   {Zacharias} M.~I.,  2013, \mn@doi [\aj]
  {10.1088/0004-6256/145/2/44}, \href
  {http://adsabs.harvard.edu/abs/2013AJ....145...44Z} {145, 44}

\makeatother
\end{thebibliography}








\bsp	
\label{lastpage}
\end{document}